\documentclass[superscriptaddress,twocolumn,aps,pra,preprintnumbers,notitlepage]{revtex4-1}
\usepackage{amsmath,amssymb,esint}
\usepackage{lipsum} %\begin{widetext} \end{widetext}
\usepackage[latin9]{inputenc}
\usepackage{graphicx}% Include figure files
\usepackage{dcolumn}% Align table columns on decimal point
\usepackage{bbold}
\usepackage{bm}% bold math
\usepackage[usenames]{xcolor}
\usepackage[colorlinks=true,linktocpage,bookmarks=true,citecolor=blue,linkcolor=blue,urlcolor=blue,hyperfootnotes=true,pageanchor=true]{hyperref}
\usepackage[caption=false,labelformat=simple]{subfig}
\usepackage{accents}
\usepackage[export]{adjustbox}
\usepackage{mathtools}
\usepackage{float}

\begin{document}

\title{Odd-frequency pair density wave in the Kitaev-Kondo lattice model}

\author{Vanuildo S. de Carvalho}
\affiliation{Instituto de F\'{i}sica Gleb Wataghin, University of Campinas (Unicamp), 13083-859, Campinas-SP, Brazil}
\affiliation{Instituto de F\'{i}sica, Universidade Federal de Goi\'as, 74.001-970, Goi\^ania-GO, Brazil}

\author{Rafael M. P. Teixeira}
\affiliation{Instituto de F\'{i}sica, Universidade Federal de Goi\'as, 74.001-970, Goi\^ania-GO, Brazil}

\author{Hermann Freire}
\affiliation{Instituto de F\'{i}sica, Universidade Federal de Goi\'as, 74.001-970, Goi\^ania-GO, Brazil}

\author{Eduardo Miranda}
\affiliation{Instituto de F\'{i}sica Gleb Wataghin, University of Campinas (Unicamp), 13083-859, Campinas-SP, Brazil}

\date{\today}

\begin{abstract}
We investigate the properties of the Kitaev-Kondo lattice model defined on a bilayer honeycomb lattice by means of the SO(3) Majorana representation for spin-$1/2$ moments. We first consider the pairing of neighboring sites for the parent Kitaev spin liquid (KSL) Hamiltonian to render the Majorana and the spin-$1/2$ Hilbert spaces perfectly equivalent to each other. As a consequence, we demonstrate that this decoupling of the Kitaev interaction in terms of the SO(3) Majorana fermions reproduces exactly the spectrum of the KSL model alone. Then, by considering the effect of a local Kondo coupling $J_K$ in the model and decoupling it in terms of an order parameter that physically must have a finite staggering phase, we obtain that the system undergoes a quantum phase transition from a fractionalized Fermi liquid to a nematic triplet superconducting (SC) phase as $J_K$ is increased. Depending on the model parameters, this SC phase can exhibit either Dirac points, Bogoliubov-Fermi lines, or Bogoliubov-Fermi surfaces as nodal bulk manifolds. The surface states in this latter case are also characterized by topologically protected antichiral edge modes. The SC phase breaks time-reversal symmetry and exhibits a coexistence of a dominant odd-frequency pairing with a small even-frequency component for electronic excitations localized on sites of the same sublattice of the system. Finally, we show that this SC phase is in fact a pair-density-wave state, with Cooper pairs possessing  a finite center-of-mass momentum in zero magnetic field.
\end{abstract}

\maketitle

\section{Introduction}

Strongly correlated electronic models are at the forefront of research in condensed matter physics, since they are known to host novel emergent phases ranging from various types of non-Fermi liquid states to high-temperature superconductivity observed in many systems. Despite this fundamental importance, solving exactly those models remains a rather intricate task in general (except for a few important cases) and those exact solutions provide crucial insights and motivate the investigation of modified strongly correlated models that include additional ingredients, which may describe more accurately the experimental situation. 

The Kitaev-Kondo lattice model is an important recent example of a strongly correlated model that cannot be solved exactly: It refers to a bilayer system with hexagonal lattices constituted by itinerant electrons in one layer and spin-$1/2$ magnetic moments in the other layer; the latter are described in terms of Kitaev exchange-frustrated interactions \cite{Kitaev-AP(2006),Trebst-PRB(2016)}. Those two subsystems interact via a local Kondo exchange interaction, which couples the magnetic moments of the parent Kitaev spin liquid (KSL) to the spin of the conduction electrons \cite{Vojta-PRB(2018),Kim-PRB(2018)}. Since the density of states of the parent KSL is linearly dependent on the energy, the Kondo interaction is thus a perturbatively irrelevant operator in the renormalization-group sense. As a result, the quasi-particle excitations of the KSL and the conduction electrons stay decoupled for small values of the Kondo interaction \cite{Fradkin-PRL(1990), Ingersent-PRB(1998), Vojta-PRB(2004), Vojta-JPCM(2005), Vojta-PM(2006)}, forming an exotic state known as fractionalized Fermi liquid (FL*) \cite{Senthil-PRL(2003),Senthil-PRB(2004)}, which might be eventually relevant to unveil the underlying mechanism of high-temperature superconductivity. In addition, only the conduction electrons contribute to the Fermi surface volume in the FL* state, which leads to the violation of the Luttinger's theorem \cite{Oshikawa-PRL(2000)}. 

The first attempt in the literature to address the properties of the Kitaev-Kondo lattice model has appeared only recently \cite{Vojta-PRB(2018)}. By means of an Abrikosov-like mean-field analysis, those authors showed that within this approximation the Kitaev-Kondo lattice model undergoes a first-order quantum phase transition (QPT) from a FL* to a superconducting (SC) phase. According to their results, this latter phase displays Bogoliubov-point nodes in the excitation spectrum, triplet pairing between nearest-neighbor sites and is characterized by the breaking of the $C_3$ lattice symmetry with the consequent emergence of nematic order. In addition, with further increase of the Kondo interaction, the system might exhibit another QPT to a conventional metallic phase featuring well-defined quasi-particle excitations, commonly referred to as a heavy Fermi liquid. Moreover, another very recent mean-field study of this model \cite{Kim-PRB(2018)}, which extended the previous analysis by including all 92 possible mean-field channels, indicated a possible existence of a QPT from the FL* to a fully-gapped non-trivial topological SC state. More specifically, in this latter scenario, the system would initially undergo a first-order QPT from a FL* to a topological SC phase with triplet pairing correlations belonging to Class D (which breaks time-reversal symmetry) in the ten-fold way classification scheme of Ref. \cite{Ludwig-PRB(2008)}. For even larger values of the Kondo interaction, those authors found an additional second-order QPT to a time-reversal symmetric topological SC phase (classified as Class DIII \cite{Ludwig-PRB(2008)}), which also displays triplet pairing correlations. 

\begin{figure}[t]
\centering
\includegraphics[width=0.8\linewidth]{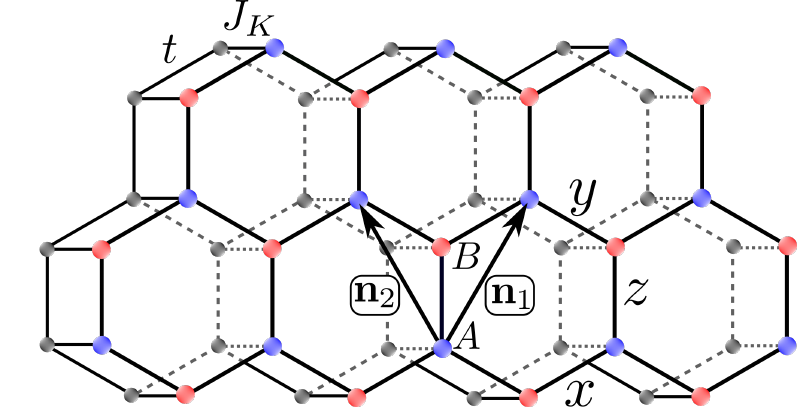}
\caption{Schematic representation of the bilayer honeycomb lattice containing the spin-$1/2$ (upper plane) and the conduction-electron (lower plane) degrees of freedom of the Kitaev-Kondo lattice model. The spin-$1/2$ magnet moments interact among themselves via the nearest-neighbor Kitaev interactions $K_{x,y,z}$, which are indicated here by $x$, $y$ and $z$. The conduction electrons are described by the hopping term $t$ and chemical potential $\mu$ that account for their kinetic energy. The interaction between these two subsystems is given by the Kondo exchange coupling $J_K$. Here, $A$ and $B$ stand for the sites of two independent sublattices, whereas $\mathbf{n}_{1}=\frac{\sqrt{3}a}{2}(1,\sqrt{3})$, $\mathbf{n}_{2}=\frac{\sqrt{3}a}{2}(-1,\sqrt{3})$ refer to the lattice vectors.}\label{Honeycomb_Lattice}
\end{figure}

Given the increasing interest in this important topic in recent years, we will revisit the Kitaev-Kondo lattice model from a different perspective in the present work. Instead of using the Abrikosov-like mean-field analysis considered in previous works, we employ here from the SO(3) Majorana fermion representation \cite{Martin-PRSL(1959), Tsvelik-PRL(1992), Sachdev-PRB(2011)} in order to reproduce the exact solution of the Kitaev model on the honeycomb lattice (a discussion about the different representations to solve this model can be found, e.g., in Ref. \cite{Perkins-PRB(2018)}). The SO(3) Majorana fermion representation will allow us to perform an exact mapping of the spin-$1/2$ Hilbert space onto the Majorana Hilbert space, when the pairing of neighboring sites is employed along one of the three bond directions of the honeycomb lattice. More importantly, one can also demonstrate that this procedure leads to an alternative solution of the Kitaev model \cite{Perkins-PRB(2018)}, which has the advantage of avoiding unphysical states in the model. Moreover, as will become clear shortly, it is mathematically appealing from the point of view of the SO(3) representation to decouple the Kondo interaction in terms of the Coleman-Miranda-Tsvelik (CMT) order parameter that pairs a conduction electron to a Majorana fermion, which was first proposed in the context of heavy-fermion superconductivity \cite{Coleman-PB(1993), Coleman-PRL(1993),Coleman-PRB(1994), Baskaran-ArXiV(2015), Tsvelik-PRL(2017)}. This approach will naturally give rise to a SC state with odd-frequency pairing \cite{Berezinskii-JETP(1974), Kirkpatrick-PRL(1991), Balatsky-PRB(1992), Belitz-PRB(1999), Tanaka-JPSJ(2012), Balatsky-RMP(2020), Pereira-PRB(2020), Miranda-PRR(2020), Cayao-EPJT(2020), Chakraborty-NJP(2021)}, whose stability is rooted in the description of the CMT order parameter in terms of a staggered phase \footnote{In the original CMT model \cite{Coleman-PRL(1993),Coleman-PRB(1994)}, the finite staggered phase of the order parameter describing the pairing of a conduction electron to a Majorana fermion is a necessary condition to obtain a positive-definite Meissner stiffness.}. As a consequence, the center of mass of the Cooper pairs has a finite momentum in zero magnetic field, making the SC obtained for this model an example of a pair-density-wave (PDW) state \cite{Himeda-PRL(2002), Berg-PRL(2007), Agterberg-NatPhys(2008), Berg-PRL(2010), PALee-PRX(2014), Fradkin-RMP(2015), Kloss-RPP(2016), Hamidian-Nature(2016), Venderley-SciAdv(2019), Agterberg-ARCMP(2020), Liu-arXiv(2020), Peng-arXiv(2020), Barnerjee-arXiv(2020)}. Depending on the model parameters, the system can exhibit either Dirac points, Bogoliubov-Fermi line, or Bogoliubov-Fermi surfaces \cite{Brydon-PRB(2018)} as the nodal manifolds of the bulk spectrum \footnote{Here we differentiate these three cases by defining that the Bogoliubov-Fermi surfaces have a finite enclosed area, whereas the Dirac nodes and the Bogoliubov-Fermi lines have zero enclosed area.}. For the latter case, the system also displays topologically protected antichiral modes as surface states \cite{Franz-PRL(2018)}. In addition, all SC phases also break both the inversion and the $C_3$ rotational symmetry related to the interchange of bonds and spin operators. 

We point out that a SC state with odd-frequency pairing component was overlooked in the mean-field approaches performed in the previous studies of the Kitaev-Kondo lattice model, since those works only considered the possibility of a SC order parameter defined for fermion fields at different sublattices. As we will demonstrate below, the odd-frequency pairing component appears when one considers the SC order parameter defined for fermion fields at the same sublattice in the model. Moreover, we emphasize here that it is necessary to take into account the staggering phase of the order parameter associated with the pairing of the conduction electrons to the Majorana fermions in the KSL. Due to this fact, we will show here that the resulting SC states have always the lowest ground-state energy in the present model.

On the experimental front, one of the motivations for studying the Kitaev-Kondo lattice model was provided recently with the observation that the Kitaev material $\alpha$-RuCl$_3$ unexpectedly starts to conduct electrical current when placed in contact with monolayer graphene \cite{Henriksen-PRB(2019)}, due to proximity effects. In addition, one can also envision the Kitaev-Kondo lattice heterostructure as another possible alternative platform to the celebrated twisted bilayer graphene \cite{Cao-Nature(2018a),Cao-Nature(2018b)}, in order to investigate the effects of strong correlations on the conduction electrons due to the proximity of a Mott insulator. 

This paper is organized as follows. In Sec. \ref{Sec_II}, we describe the Kitaev-Kondo lattice model and express the spin-$1/2$ magnetic moment operators in terms of Majorana fermions using the SO(3) representation. Next, in Sec. \ref{Sec_III}, we introduce the mean-field order parameters used to decouple the Kitaev and Kondo interactions. The former decoupling is done in the spin-liquid and magnetic channels, while the latter is performed in terms of the CMT order parameter. The solution of the mean-field equations is then described in Sec. \ref{Sec_IV}. In Sec. \ref{Sec_V}, we address the properties of the SC states that emerge out of the FL* phase and present the ground-state phase diagram of the model as a function of the electron doping, hopping parameter, and the Kondo interaction. Finally, in Sec. \ref{Sec_VI}, we present our conclusions. Technical details of our mean-field calculations are left to the Appendices.

\section{Model}\label{Sec_II}

To begin with, we write down the Hamiltonian of the Kitaev-Kondo lattice model as $H=H_t+H_K+H_J$, where
\begin{align}
H_t = &-t\sum\limits^{}_{\langle j,k\rangle,\sigma}(\psi^{\dagger}_{j,\sigma}\psi_{k,\sigma}+\text{H.c.})-\mu\sum\limits^{}_{j,\sigma}\psi^{\dagger}_{j,\sigma}\psi_{j,\sigma},\label{Ham_Psi}\\
H_K = &-\sum_{\alpha=x,y,z}\sum\limits^{}_{\langle j,k\rangle_{\alpha}}K_{\alpha}S^{\alpha}_{j}S^{\alpha}_{k},\label{Ham_Kitaev}\\
H_J = &\ J_{K}\sum\limits^{}_{j,\sigma,\sigma'}(\psi^{\dagger}_{j,\sigma}\boldsymbol{\sigma}_{\sigma,\sigma'}\psi_{j,\sigma'})\cdot\mathbf{S}_{j},\label{Ham_Kondo}
\end{align}
where $H_t$ is the tight-binding Hamiltonian of the conduction electrons with hopping $t$ and chemical potential $\mu$; $\psi^{\dagger}_{j,\sigma}$ ($\psi_{j,\sigma}$) are the creation (annihilation) operators for conduction electrons with spin projection $\sigma \in \{\uparrow, \downarrow \}$, which are located on the site $j$ of the honeycomb lattice (see Fig. \ref{Honeycomb_Lattice}). $H_K$ refers to the Hamiltonian of the parent KSL, which is described in terms of three frustrated exchange interactions $K_\alpha$ ($\alpha\in\{x,y,z\}$) involving the spin-$1/2$ moments given by $\widetilde{\mathcal{S}}^\alpha_j \equiv S^\alpha_j/2$. Note that, in this convention, the operators $S^\alpha_j$ that will be used throughout this work are defined as twice the operators corresponding to the spin-$1/2$ moments; this is done to simplify the expressions obtained below. Lastly, $H_J$ denotes the Kondo Hamiltonian defined in terms of the local exchange coupling $J_K$ between the conduction electron spin and the localized spin-$1/2$ moments. 

In order to map out the phase diagram of $H$, we make use of the SO(3) Majorana representation for spin-$1/2$ moments. It is given by \cite{Martin-PRSL(1959), Tsvelik-PRL(1992), Sachdev-PRB(2011)}:
\begin{equation}\label{Eq_SpinMajorana}
S^{\alpha}_{j} = - \frac{i}{2} \epsilon^{\alpha\beta\gamma}c^{\beta}_{j}c^{\gamma}_{j},
\end{equation}
where $c^{\alpha}_{j}$ are Majorana fermion operators, which, by definition, obey $c^{\alpha\dagger}_{j}=c^{\alpha}_{j}$ and the anti-commutation relations $\{c^{\alpha}_{j},c^{\beta}_{k}\}=2\delta_{\alpha\beta}\delta_{jk}$ \cite{Tsvelik-PRL(1992),Coleman-PRB(1994)}. The SO(3) Majorana representation reproduces both the SU(2) and Clifford algebras associated with the spin-$1/2$ operators. However, it is overcomplete because the dimension ($\operatorname{dim}$) of the Majorana Hilbert space $\mathcal{H}_M$ for a system with, say, an even number $N$ of spins is $\operatorname{dim}(\mathcal{H}_M)=2^{3N/2}$, while the dimension of the spin Hilbert space $\mathcal{H}_S$ is $\operatorname{dim}(\mathcal{H}_S)=2^N$. In order to explain why there is a difference in the dimension of $\mathcal{H}_M$ and $\mathcal{H}_S$, we follow Ref. \cite{Sachdev-PRB(2011)} and define the operators
\begin{align}
\hat{\mathcal{O}}_j&\equiv ic^x_jc^y_jc^z_j,\label{Eq_OOp}\\
\hat{\mathcal{T}}_{jk}&\equiv\hat{\mathcal{O}}_j\hat{\mathcal{O}}_k.\label{Eq_TOp}
\end{align}
First, we note that $\hat{\mathcal{O}}_j$ commutes with all the spin-$1/2$ operators $S^\alpha_k$, i.e., $[\hat{\mathcal{O}}_j,S^\alpha_k]=0$. Consequently, we also have the commutation relations $[\hat{\mathcal{T}}_{jk},S^\alpha_l]=0$. Due to the property $\hat{\mathcal{T}}^2_{jk}=-1$, one can also demonstrate that the eigenvalues of the operators $\hat{\mathcal{T}}_{jk}$ are simply given by $\mathcal{T}_{jk}=\pm i$. If we now perform the pairing of sites along, for example, the $z$ bonds of the honeycomb lattice (see Fig. \ref{Honeycomb_Lattice}), so that the first and second indices in the subscript of $\mathcal{T}_{jk}$ refer, respectively, to the sites of the $A$ and $B$ sublattices, we obtain that there are $2^{N/2}$ possibilities to cover the honeycomb lattice by choosing different values of $\mathcal{T}_{jk}=\pm i$. This is essentially the ratio $\operatorname{dim}(\mathcal{H}_M)/\operatorname{dim}(\mathcal{H}_S)$ between the Majorana and spin Hilbert space. Hence, an exact mapping between the two Hilbert spaces $\mathcal{H}_M$ and $\mathcal{H}_S$ can be achieved by choosing the same eigenvalue of $\hat{\mathcal{T}}_{jk}$ for the set of nearest-neighbor sites $\langle j,k\rangle_z$ along a $z$ bond. Indeed, to rigorously demonstrate this proposition, we first define the number operator $\hat{n}^\alpha_{jk}\equiv f^{\alpha\dagger}_{jk}f^{\alpha}_{jk}$, where 
\begin{equation}
f^\alpha_{jk}\equiv\frac{1}{2}(c^\alpha_j+ic^\alpha_k)
\end{equation}
are complex fermion operators for nearest-neighbor sites $j$ and $k$ along a $z$ bond. In terms of the SO(3) Majorana fermions, the number operator reads
\begin{equation}\label{Eq_NumOp}
\hat{n}^\alpha_{jk}=\frac{1}{2}(1+ic^\alpha_jc^\alpha_k).
\end{equation}
The substitution of Eq. \eqref{Eq_NumOp} into Eq. \eqref{Eq_TOp} yields
\begin{equation}\label{Eq_TNOp}
\hat{\mathcal{T}}_{jk}=-i(-1)^{n^x_{jk}+n^y_{jk}+n^z_{jk}},
\end{equation}
where $n^\alpha_{jk}\in\{0,1\}$ are the eigenvalues of $\hat{n}^\alpha_{jk}$. Next, we arrange the vectors of state associated with a pair of nearest-neighbor sites $\langle j,k\rangle_z$ within $\mathcal{H}_M$ as 
\begin{equation}\label{Eq_VecHilbert}
|n^x_{jk},n^y_{jk},n^z_{jk}\rangle=(f^{x\dagger}_{jk})^{n^x_{jk}}(f^{y\dagger}_{jk})^{n^y_{jk}}(f^{z\dagger}_{jk})^{n^z_{jk}}|0,0,0\rangle.
\end{equation}
According to Eqs. \eqref{Eq_TNOp} and \eqref{Eq_VecHilbert}, the Majorana Hilbert space defined by $\mathcal{T}_{jk}=-i$ for every pair of nearest-neighbor sites $\langle j,k\rangle_z$ is spanned by the basis $\{|0,0,0\rangle,|1,1,0\rangle,|1,0,1\rangle,|0,1,1\rangle\}$. Consequently, this basis can be used to express the vectors of the spin-$1/2$ basis $\{|\uparrow,\uparrow\rangle,|\uparrow,\downarrow\rangle,|\downarrow,\uparrow\rangle,|\downarrow,\downarrow\rangle\}$ for the same pair of sites $\langle j,k\rangle_z$, resulting in a bijective mapping between the two bases and, therefore, in a perfect match between the dimensions of both Hilbert spaces for the spin and Majorana degrees of freedom.

\begin{figure*}[t]
\centering \includegraphics[width=0.43\linewidth,valign=t]{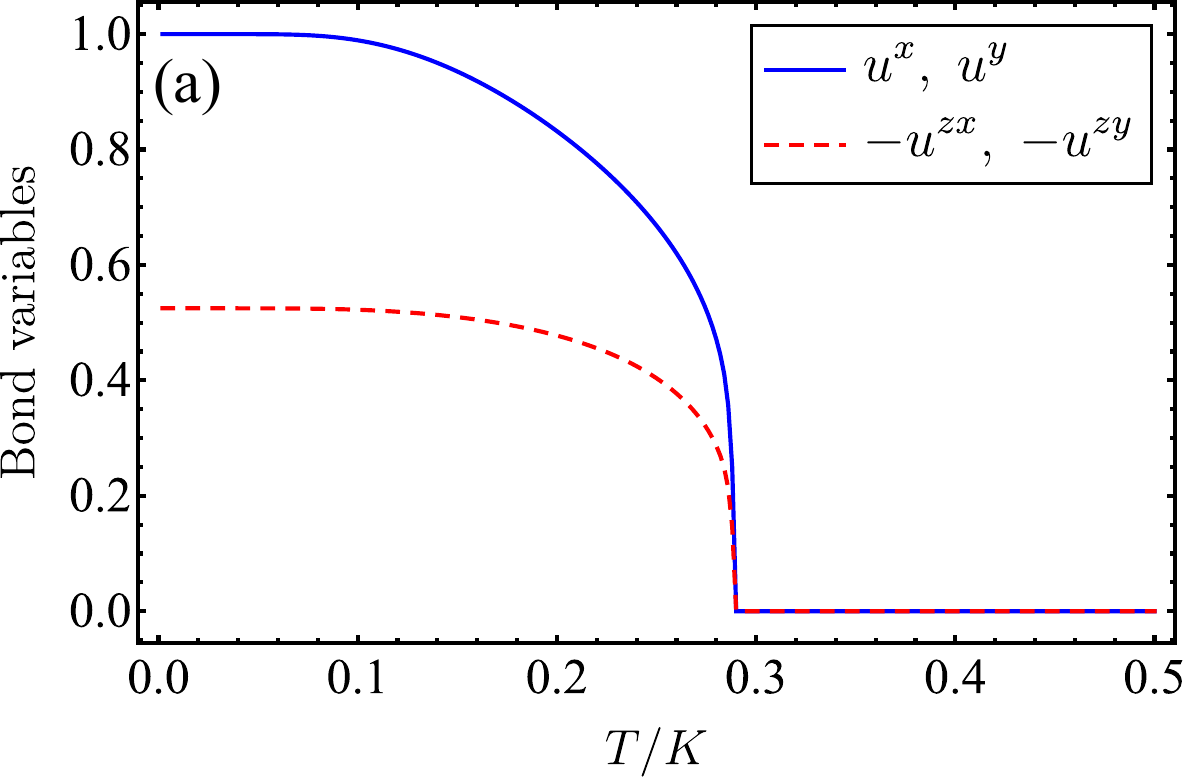}\hfil{}\includegraphics[width=0.445\linewidth,valign=t]{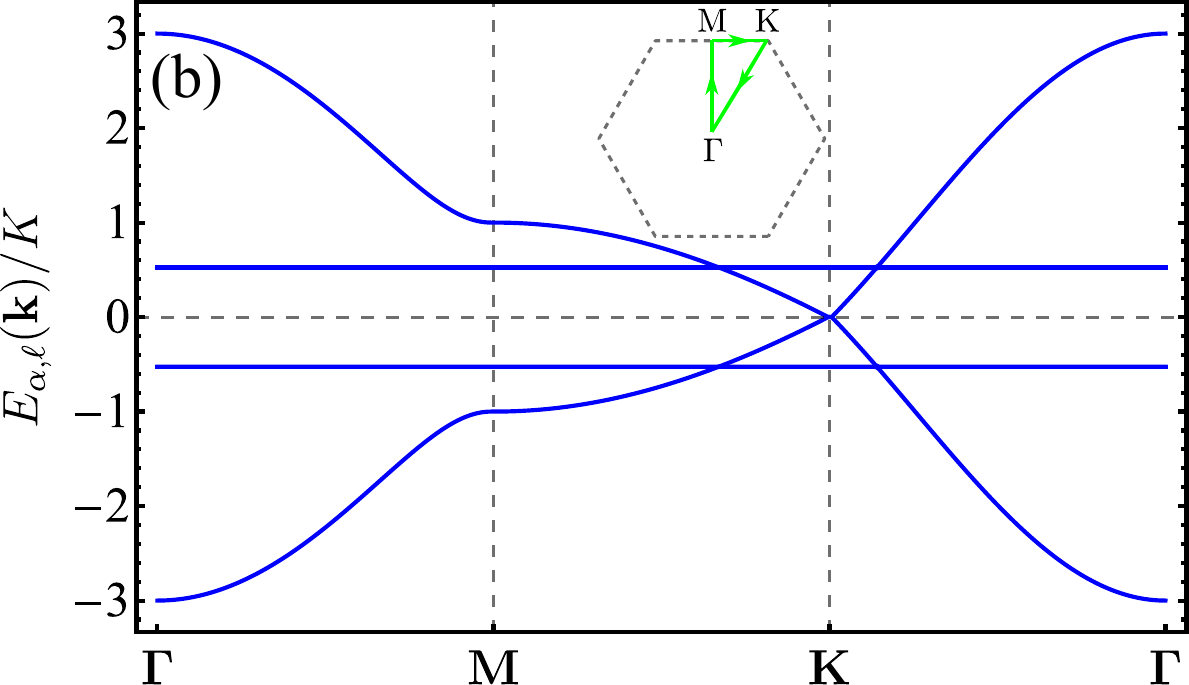}
\caption{(a) Behavior of the order parameter of the Kitaev model as a function temperature for isotropic exchange interactions. (b) Energy dispersions of the mean-field KSL Hamiltonian $H_{\text{\tiny MF},K}$ for the parameters obtained in Eq. \eqref{Eq_Sol_OP}. The dispersionless bands shown here are doubly degenerate. The other two bands are gapless and exhibit a Dirac node at momentum $\mathbf{K}=\frac{2\pi}{3a}\left(\frac{1}{\sqrt{3}},1\right)$. The inset shows the first Brillouin zone of the honeycomb lattice with a path (thick green line) going through its highly symmetric points.}\label{Kitaev_MF_Sol_Dispersions}
\end{figure*}

We can now apply these findings to the Kitaev-Kondo lattice model. First, by inserting Eqs. \eqref{Eq_SpinMajorana}--\eqref{Eq_TOp} into Eq. \eqref{Ham_Kitaev}, one obtains that the KSL Hamiltonian $\mathcal{H}_K$ becomes \cite{Perkins-PRB(2018)}:
\begin{eqnarray}
H_{K}&=&\; \sum\limits^{}_{\langle j,k\rangle_{x}}K_{x}(ic^{y}_{j}c^{y}_{k})ic^{z}_{j}c^{z}_{k}+\sum\limits^{}_{\langle j,k\rangle_{y}}K_{y}(ic^{x}_{j}c^{x}_{k})ic^{z}_{j}c^{z}_{k}\nonumber\\
&+&\sum\limits^{}_{\langle j,k\rangle_{z}}K_{z}\hat{\mathcal{T}}_{jk}c^{z}_{j}c^{z}_{k}.
\end{eqnarray}
Since the operators $\hat{\mathcal{T}}_{jk}$ are defined here just for the pair of sites $\langle j,k\rangle_z$, we get $[\hat{\mathcal{T}}_{jk},H_K]=0$. Moreover, one can also check that $[\hat{\mathcal{T}}_{jk},H_t]=[\hat{\mathcal{T}}_{jk},H_J]=0$, which consequently implies that $\hat{\mathcal{T}}_{jk}$ is a constant of motion of $H$ and thus can be substituted by its eigenvalues $\mathcal{T}_{jk}=\pm i$. By taking into account these facts and then defining the spinor $\psi_{j}\equiv(\psi_{j,\uparrow},\psi_{j,\downarrow})^{T}$, the terms in the definition of $H$ become
\begin{align}
H_{t}=&-t\sum\limits^{}_{\langle j,k\rangle}(\psi^{\dagger}_{j}\psi_{k}+\text{H.c.})-\mu\sum\limits^{}_{j}\psi^{\dagger}_{j}\psi_{j},\label{Eq_HamPsi}\\
H_{K}=&\; \sum\limits^{}_{\langle j,k\rangle_{x}}K_{x}(ic^{y}_{j}c^{y}_{k})ic^{z}_{j}c^{z}_{k}+\sum\limits^{}_{\langle j,k\rangle_{y}}K_{y}(ic^{x}_{j}c^{x}_{k})ic^{z}_{j}c^{z}_{k}\nonumber\\
&+\sum\limits^{}_{\langle j,k\rangle_{z}}K_{z}\mathcal{T}_{jk}c^{z}_{j}c^{z}_{k},\label{Eq_HamK}\\
H_{J}=&-\frac{J_{K}}{2}\sum\limits^{}_{j}\psi^{\dagger}_{j}[i\boldsymbol{\sigma}\cdot(\mathbf{c}_{j}\times\mathbf{c}_{j})]\psi_{j},\label{Eq_HamJ}
\end{align}
where we are considering $\mathcal{T}_{jk}=+i\; (-i)$ for a site $j$ in the $A\; (B)$ sublattice. As explained earlier, this choice of eigenvalues for the operators $\hat{\mathcal{T}}_{jk}$ allows us to establish a one-to-one correspondence between the bases of the Majorana and spin-$1/2$ Hilbert spaces. In addition, it also gives the lowest free energy for the KSL sector of the model, when the Kondo interaction $J_K$ is set to zero \cite{Perkins-PRB(2018)}.

\section{Mean-field theory of the Kitaev-Kondo lattice model}\label{Sec_III}

As mentioned previously, the Kitaev-Kondo lattice model given by Eqs. \eqref{Eq_HamPsi}--\eqref{Eq_HamJ} is not exactly solvable and, due to this fact, we will proceed here with a mean-field analysis of its low-energy properties. The mean-field decoupling of the Kitaev Hamiltonian in both the spin-liquid and magnetic channel yields
\begin{align}
H_{\text{\tiny MF},K} & = \sum_{\langle j,k\rangle_x}K_x[\langle ic^y_jc^y_k\rangle ic^z_jc^z_k+ic^y_jc^y_k\langle ic^z_jc^z_k\rangle\nonumber\\&-\langle ic^y_jc^y_k\rangle\langle ic^z_jc^z_k\rangle]\nonumber\\
& + \sum_{\langle j,k\rangle_y}K_y[\langle ic^x_jc^x_k\rangle ic^z_jc^z_k+ic^x_jc^x_k\langle ic^z_jc^z_k\rangle\nonumber\\&-\langle ic^x_jc^x_k\rangle\langle ic^z_jc^z_k\rangle]\nonumber\\
& - \sum_{\langle j,k\rangle_x}K_x[\langle ic^y_jc^z_j\rangle ic^y_kc^z_k+ic^y_jc^z_j\langle ic^y_kc^z_k\rangle\nonumber\\&-\langle ic^y_jc^z_j\rangle\langle ic^y_kc^z_k\rangle]\nonumber\\
& - \sum_{\langle j,k\rangle_y}K_y[\langle ic^z_jc^x_j\rangle ic^z_kc^x_k+ic^z_jc^x_j\langle ic^z_kc^x_k\rangle\nonumber\\&-\langle ic^z_jc^x_j\rangle\langle ic^z_kc^x_k\rangle]\nonumber\\
& + \sum\limits^{}_{\langle j,k\rangle_{z}}K_{z}\mathcal{T}_{jk}c^{z}_{j}c^{z}_{k}. \label{Eq_MFHamK}
\end{align}
On the one hand, the spin-liquid degrees of freedom of this Hamiltonian refer to the bond variables $u^\alpha_{\langle j,k\rangle_\beta}\equiv\langle ic^\alpha_jc^\alpha_k\rangle$, which are the nearest-neighbor mean-field order parameters associated with the Majorana fermion species $\alpha\in\{x,y,z\}$ along the bond $\langle j,k\rangle_\beta$. For the case of a homogeneous KSL system, these order parameters simplify to
\begin{align}
u^x_{\langle j,k\rangle_y}&=-u^x_{\langle k,j\rangle_y}=u^x,\\
u^y_{\langle j,k\rangle_x}&=-u^y_{\langle k,j\rangle_x}=u^y,\\
u^z_{\langle j,k\rangle_x}&=-u^z_{\langle k,j\rangle_x}=u^{zx},\\
u^z_{\langle j,k\rangle_y}&=-u^z_{\langle k,j\rangle_y}=u^{zy}.
\end{align}
On the other hand, the fluctuations in the magnetic channel are described by the two order parameters $m^x_j \equiv \langle i c^z_j c^y_j \rangle$ and $m^y_j \equiv \langle i c^x_j c^z_j \rangle$, which correspond to the expectation values $\langle S^x_j \rangle$ and $\langle S^y_j \rangle$ of the $\hat{\mathbf{x}}$ and $\hat{\mathbf{y}}$ components of the in-plane magnetization, respectively. We will consider in what follows the variation of $m^x_j$ and $m^y_j$ with respect to the sublattice indices, which leads to four order parameters $m^x_{A (B)}$ and $m^y_{A (B)}$.

The mean-field solution of the Kitaev Hamiltonian in Eq. \eqref{Eq_MFHamK} shows that $m^x_{A (B)}$ and $m^y_{A (B)}$ are zero for all temperatures, which is consistent with the exact solution of this model \cite{Kitaev-AP(2006)}. In contrast, the bond variables turn out to be finite as the temperature is lowered. Indeed, the solution of the mean-field equation for isotropic Kitaev interactions $K_x = K_y = K_z = K$ shows the existence of a critical temperature $T_c$, below which all bond variables become non-zero [see Fig. \ref{Kitaev_MF_Sol_Dispersions}(a)]. In the $T\rightarrow 0$ limit, they approach the constant values
\begin{equation}\label{Eq_Sol_OP}
u^x_0 = 1, \;\; u^y_0 = 1, \;\; u^{zx}_0 = - 0.524864, \;\; u^{zy}_0 = - 0.524864.
\end{equation}
We note that this solution is equivalent to the one found in Ref. \cite{Kim-PRB(2018)} for the same model, but by making use of the Kitaev Majorana representation for the spin-$1/2$ moments. In this case, it was employed four Majorana fermions to represent the spin-$1/2$ algebra and a projection operator to define the physical states of the model. Moreover, as shown in Fig. \ref{Kitaev_MF_Sol_Dispersions}(b), the spectrum of the Majorana fermions obtained from the solution in Eq. \eqref{Eq_Sol_OP} exhibits two bands with gapless excitations and two others that do not disperse. Besides, the latter bands turn out to be doubly degenerate.  

In addition, by substituting the values of the order parameters in Eq. \eqref{Eq_Sol_OP}, as well as $m^x_{A (B)} = 0$ and $m^y_{A (B)} = 0$, into the KSL free energy and then taking its $T\rightarrow 0$ limit, we find that the mean-field ground state energy per unit cell evaluates to $E^{\text{\tiny (0)}}_{\text{\tiny MF},K} = - 1.5746 K$. It turns out that the value of $E^{\text{\tiny (0)}}_{\text{\tiny MF},K}$ obtained by the mean-field theory using the SO(3) Majorana spin representation is equal to the exact ground state energy of the KSL model found in Ref. \cite{Kim-PRB(2018)}.

Regarding the Kondo Hamiltonian $H_J$ written in terms of the SO(3) Majorana fermion operators [see Eq. \eqref{Eq_HamJ}], it is interesting to note that this is formally identical to the expression that appears in a model for heavy-fermion compounds known as Coleman-Miranda-Tsvelik (CMT) model \cite{Coleman-PRL(1993), Coleman-PRB(1994)}. One of the main features of the latter model is that it exhibits a triplet superconducting phase with odd-frequency pairing \cite{Maciejko-PRB(2017), Balatsky-RMP(2020)}. In the present context, motivated by the investigation of a possible emergence of such a state also in the Kitaev-Kondo lattice model, we will proceed by employing here the same strategy as devised in Refs. \cite{Coleman-PRL(1993),Coleman-PRB(1994)} for treating the Kondo interaction. Therefore, we rewrite $H_J$ as
\begin{equation}
H_J=-\frac{J_{K}}{2}\sum\limits^{}_{j}\psi^{\dagger}_{j}(\boldsymbol{\sigma}\cdot\mathbf{c}_{j})^2\psi_{j}+\frac{3J_{K}}{4}\sum\limits^{}_{j}\psi^{\dagger}_{j}\psi_{j},
\end{equation}
where we have made use of the identity $i\boldsymbol{\sigma}\cdot(\mathbf{c}_j\times\mathbf{c}_j)=(\mathbf{c}_j\cdot\boldsymbol{\sigma})^2-3/2$. The second term on the right-hand side is not of great importance in the analysis of the model, since it can be absorbed into a redefinition of the chemical potential. On the other hand, the instabilities of the system induced by the Kondo interaction are described by the first term in that equation. With that in mind, this latter Hamiltonian can be simplified according to the mean-field decoupling
\begin{equation}\label{Eq_MF_HamJ_01}
H_{\text{\tiny MF},J}=\sum\limits^{}_j[\psi^{\dagger}_j(\boldsymbol{\sigma}\cdot\mathbf{c}_j)V_j+V^{\dagger}_j(\mathbf{c}_j\cdot\boldsymbol{\sigma})\psi_j]+2\sum\limits^{}_j\frac{|V_j|^2}{J_K},
\end{equation}
where the CMT order parameter $V_j$ is given by
\begin{equation}\label{Eq_OP_V}
V_j=\begin{pmatrix} V_{j,\uparrow} \\ V_{j,\downarrow} \end{pmatrix}=-\frac{J_K}{2}\langle(\boldsymbol{\sigma}\cdot\mathbf{c}_j)\psi_j\rangle.
\end{equation}
Both components of $V_j$ have a real ($R$) and an imaginary ($I$) amplitude, i.e., $V_{j, \sigma} = V_{R, \sigma} + i V_{I, \sigma}$ with $\sigma \in \{ \uparrow, \downarrow \}$. In order to further simplify some calculations, we will employ in what follows the $4$-component Balian-Werthamer spinors
\begin{align}
\Psi_j&\equiv\begin{pmatrix} \psi_j \\ -i\sigma^y(\psi^{\dagger}_j)^T \end{pmatrix},\label{Eq_BW_Psi}\\
\mathcal{V}_j&\equiv\begin{pmatrix} V_j \\ -i\sigma^y(V^{\dagger}_j)^T \end{pmatrix}.\label{Eq_BW_V}
\end{align}

Next, by substituting Eqs. \eqref{Eq_BW_Psi} and \eqref{Eq_BW_V} into Eqs. \eqref{Eq_HamPsi} and \eqref{Eq_MF_HamJ_01}, we obtain
\begin{align}
H_{t}=&-\frac{t}{2}\sum\limits^{}_{\langle j,k\rangle}(\Psi^{\dagger}_{j} T^z\Psi_{k}+\text{H.c.})-\frac{\mu}{2}\sum\limits^{}_{j}\Psi^{\dagger}_{j} T^z\Psi_{j},
\end{align}
and
\begin{align}
H_{\text{\tiny MF},J}=&\;\frac{1}{2}\sum\limits^{}_j\{\Psi^{\dagger}_j[(\boldsymbol{\sigma}\otimes\mathbb{1})\cdot\mathbf{c}_j]\mathcal{V}_j+\mathcal{V}^{\dagger}_j[\mathbf{c}_j\cdot(\boldsymbol{\sigma}\otimes\mathbb{1})]\Psi_j\}\nonumber\\
&+\sum\limits^{}_j\frac{|\mathcal{V}_j|^2}{J_K},\label{Eq_MF_HamJ_02}
\end{align}
where
\begin{equation}
\boldsymbol{\sigma}\otimes\mathbb{1}=\begin{pmatrix} \boldsymbol{\sigma} & \mathbb{0} \\ \mathbb{0} & \boldsymbol{\sigma} \end{pmatrix},\; T^z = \begin{pmatrix} \mathbb{1} & \mathbb{0} \\ \mathbb{0} & -\mathbb{1} \end{pmatrix}.
\end{equation}

In order to describe the spatial fluctuations in the symmetry-broken phase, we employ the following set of transformations \cite{Coleman-PRB(1994)}:
\begin{align}
V_j&\stackrel{T_\theta}{\longmapsto} e^{i\theta_j}V_j,\label{Eq_Transf_01}\\
\mathcal{V}_j&\stackrel{T_\theta}{\longmapsto} e^{i\theta_j T^z}\mathcal{V}_j,\label{Eq_Transf_02}\\
\Psi_j&\stackrel{T_\theta}{\longmapsto} e^{i\theta_j T^z}\Psi_j,\label{Eq_Transf_03}
\end{align}
where $\theta_j=\mathbf{Q}\cdot\mathbf{R}_j$, with $\mathbf{Q}$ being an staggered wave vector. These transformations do not change the form of the Hamiltonian $H_{\text{\tiny MF},K}$ for the parent KSL and the Kondo interaction $H_{\text{\tiny MF},J}$. However, it changes the structure of the tight-binding Hamiltonian $H_{t}$ for the conduction electrons. In fact, the latter becomes
\begin{align}
H_{t} = & - \frac{t}{2}\sum\limits^{}_{\langle j,k\rangle}[\Psi^{\dagger}_{j}e^{-i\theta_j T^z} T^ze^{i\theta_k T^z}\Psi_{k}+\text{H.c.}] \nonumber \\
& - \frac{\mu}{2}\sum\limits^{}_{j}\Psi^{\dagger}_{j}T^z\Psi_{j}. \label{Eq_HamPsi_GT}
\end{align}
Notice that the set of transformations in Eqs. \eqref{Eq_Transf_01}--\eqref{Eq_Transf_03} is formally equivalent to writing the order parameter $V_j$ as $V_j=e^{i\mathbf{Q}\cdot\mathbf{R}_j}\begin{pmatrix} V_{j,\uparrow} \\ V_{j,\downarrow} \end{pmatrix}$, which despite its spinorial nature resembles the order parameter of an antiferromagnetic state.

\begin{figure*}[t]
\centering 
\includegraphics[width=0.32\linewidth,valign=t]{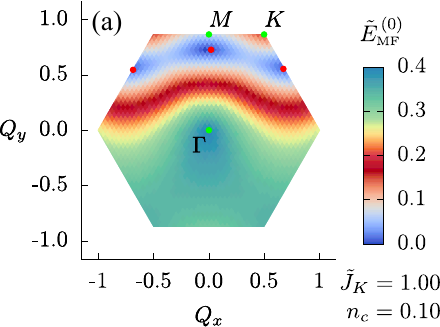} \hfil{}
\includegraphics[width=0.32\linewidth,valign=t]{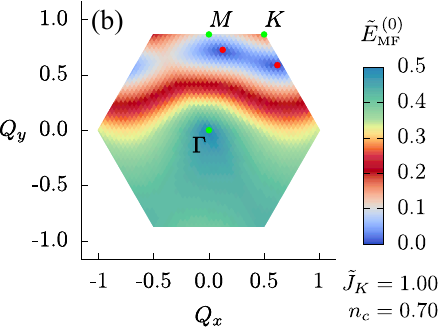} \hfil{}
\includegraphics[width=0.32\linewidth,valign=t]{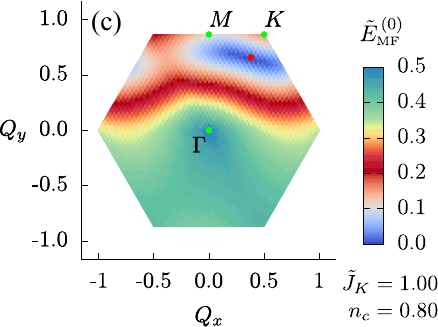} \vfil{}
\includegraphics[width=0.32\linewidth,valign=t]{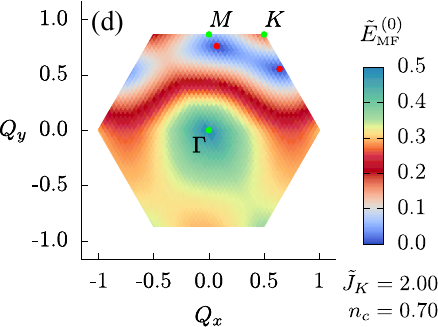} \hfil{}
\includegraphics[width=0.32\linewidth,valign=t]{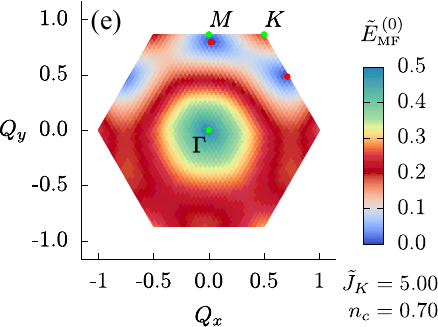} \hfil{}
\includegraphics[width=0.32\linewidth,valign=t]{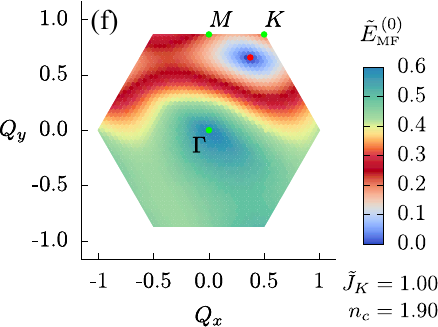} 

\caption{Dependence of the rescaled ground-state energy $\tilde{E}^{\text{\tiny (0)}}_\text{\tiny MF} = \sqrt{(E^{\text{\tiny (0)}}_\text{\tiny MF} - E^{\text{\tiny (0))}}_\text{\tiny MF,min})/K}$  on the wave vector $\mathbf{Q} = (Q_x, Q_y)$ within the first Brillouin zone of the honeycomb lattice as a function of $\tilde{J}_K=J_K/K$ and $n_c$. Here, $E^{\text{\tiny (0)}}_\text{\tiny MF,min}$ refers to the minimum value of the ground state energy. The red points mark the wave  vector $\mathbf{Q}^{*}$ that minimizes the energy. As can be seen from these density plots, in some cases, there is more than one wave vector corresponding to the same minimum value of energy within numerical accuracy. Notice that the components of $\mathbf{Q}$ are rescaled as $Q_{w}=3\sqrt{3}\left(\mathbf{Q}\cdot\hat{\mathbf{w}}\right)/(4\pi)$, where $w=x,y$.}\label{EMF_1BZ_Plots}
\end{figure*}

\section{Self-consistent solution of the mean-field equations}\label{Sec_IV}
\noindent

Before presenting the self-consistent solution of the mean-field equations (see Appendix \ref{Appendix_A}), we need to determine the staggered wave vector $\mathbf{Q}$ that gives rise to the lowest ground-state energy $E^{\text{\tiny (0)}}_\text{\tiny MF}$ in the system. The dependence of $E^{\text{\tiny (0)}}_\text{\tiny MF}$ on both the Kondo interaction and $\mathbf{Q}$ is shown in Fig. \ref{EMF_1BZ_Plots}. In  Fig. \ref{EMF_1BZ_Plots}(a), one can see that the energy is nearly symmetric with respect to the $Q_{y}$-axis and the staggered wave vector $\mathbf{Q}$ providing the lowest ground state energy $E^{\text{\tiny (0)}}_\text{\tiny MF,min}$ is near the $\mathbf{M}$ point. The other two minima correspond to values that are actually very close numerically to the first one. In fact, the difference is of the order of $10^{-4}$ in units of the Kitaev interaction $K$. Since our implementation of the numerical method for solving the coupled nonlinear equations for the order parameters has a precision of the same order, these three minimum energy values can be seen as equal within numerical error. This indicates that, for densities satisfying $n_{c}\ll 1$, the mean-field ground state energy displays a three-fold degeneracy, which is lifted as $n_{c}$ increases and turns into a two-fold degeneracy as shown in Fig. \ref{EMF_1BZ_Plots}(b), where the two degenerate minima can be seen near the $\mathbf{M}$ and $\mathbf{K}$ points for $n_c=0.7$. In Fig. \ref{EMF_1BZ_Plots}(c), this two-fold degeneracy eventually disappears and, as the conduction-electron density $n_c$ varies approximately from $0.8$ to $2.0$, the global minimum remains located approximately at the same wave vector $\mathbf{Q}^{*}$, which is both close to the $\mathbf{K}$ point and along the $\boldsymbol{\Gamma} - \mathbf{K}$ line. On the other hand, if we fix the quantity $n_c$ and then increase the Kondo interaction $J_K$ in the model, the two-fold degeneracy remains for conduction electron densities smaller than unity, as displayed in Figs. \ref{EMF_1BZ_Plots}(d) and  \ref{EMF_1BZ_Plots}(e); for larger densities, the wave vector for the ground-state minimum continues to be located at the single and unchanged wave vector $\mathbf{Q}^{*}$ mentioned previously [see Fig. \ref{EMF_1BZ_Plots}(f)]. We should also mention that the symmetries of $E^{\text{\tiny (0)}}_\text{\tiny MF}$ seen in Fig. \ref{EMF_1BZ_Plots} depend on the way one performs the pairing of neighboring sites in the Kitaev Hamiltonian to define the physical Hilbert space of the spin degrees of freedom. Because of that, $E^{\text{\tiny (0)}}_\text{\tiny MF}$ becomes rotated by an angle of $2 \pi/3$, as the pairing of sites is interchanged along the $x$, $y$, and $z$ bonds. However, the mean-field solutions resulting from this process are all degenerate to each other.

\begin{figure*}[t]
\centering 
\includegraphics[width=0.31\linewidth,valign=t]{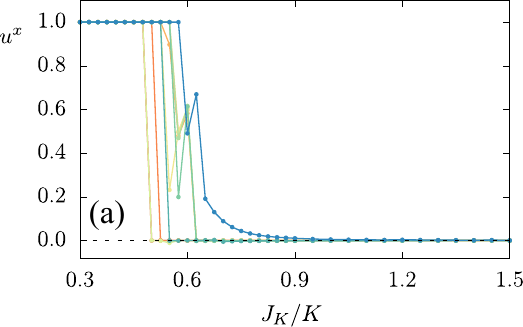} \hfil{}
\includegraphics[width=0.31\linewidth,valign=t]{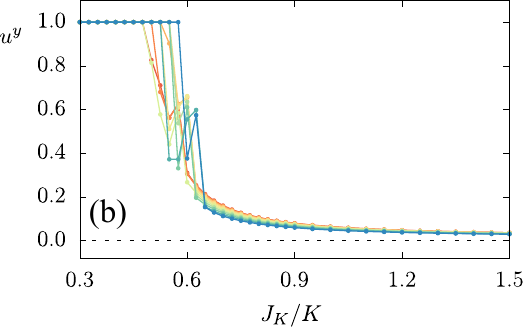} \hfil{}
\includegraphics[width=0.31\linewidth,valign=t]{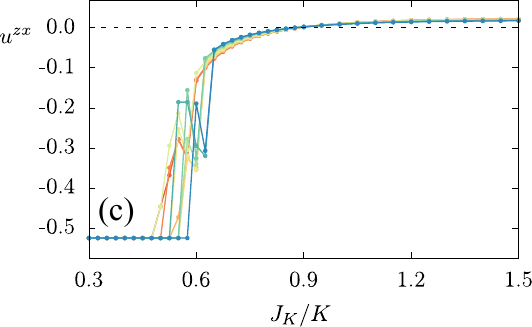} \hfil{}
\includegraphics[width=0.038\linewidth,valign=t]{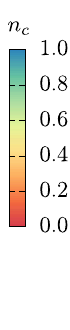} \vfil{}
\includegraphics[width=0.31\linewidth,valign=t]{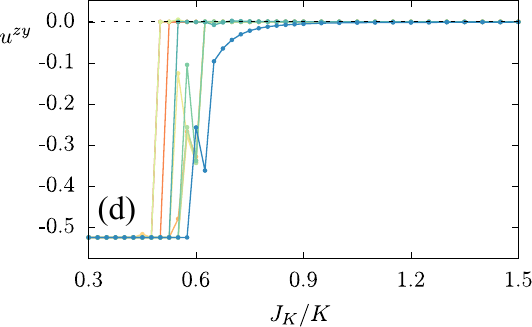} \hfil{}
\includegraphics[width=0.31\linewidth,valign=t]{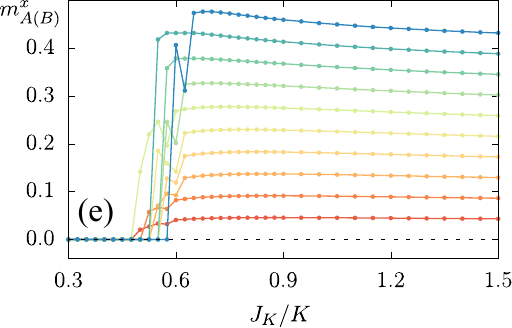} \hfil{}
\includegraphics[width=0.31\linewidth,valign=t]{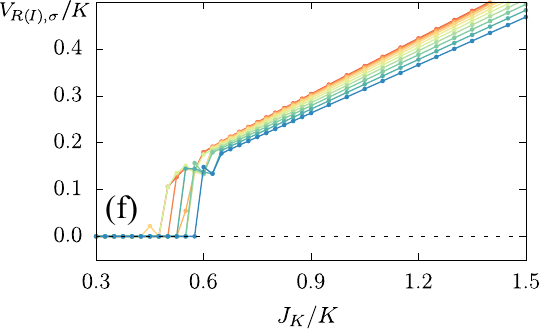} \hfil{} 
\includegraphics[width=0.038\linewidth,valign=t]{Color_Box} 

\caption{Behavior of the order parameters of the Kitaev-Kondo lattice as a function of both the Kondo interaction $J_K$ and the electron doping $n_c$, when we fix the electron hoping to $t/K = 0.2$ and $\mathbf{Q} = \mathbf{Q}^{*}$, where $\mathbf{Q}^{*}$ refers to the staggered wave vector that gives the lowest ground-state energy. (a)--(d) The bond variables $u^x$, $u^y$, $u^{zx}$, and $u^{zy}$ are strongly suppressed as the system enters the symmetry-broken phase. (e) The two components of the magnetic order parameter, $m^x_{A (B)}$ and $m^y_{A (B)}$, do not become finite at the same time as the QPT takes place. However, they are both degenerate with respect to the sublattice index. For the particular solution shown here, we find $m^y_{A (B)} = 0$. (f) The real $V_{R, \sigma}$ and imaginary $V_{I, \sigma}$ components of the CMT order parameter are found to be degenerate to each other in the present model. We point out that the non-monotonic behavior of the order parameters at the transition region is due to the numerical precision of the algorithm used to solve the mean-field equations.}\label{MFSol_Qmin}
\end{figure*}

In Fig. \ref{MFSol_Qmin}, we display the zero-temperature dependence of the mean-field order parameters on the Kondo interaction within the range $0<n_{c}\leq 1$ (the latter interval was chosen only for the sake of numerical convenience), and for the following choice of the wave vector, i.e., $\mathbf{Q} = \mathbf{Q}^{*}$. It is important to emphasize that the location of the quantum critical point separating the FL* from the symmetry-broken phase associated with a finite CMT order parameter grows monotonically with the electron doping $n_c$. As also shown in Figs. \ref{MFSol_Qmin}(a)--\ref{MFSol_Qmin}(d), the bond variables $u^x$, $u^y$, $u^{zx}$, and $u^{zy}$ are abruptly suppressed right after the system undergoes the QPT out of the FL* phase.

Regarding the self-consistent solutions of the magnetic order parameters $m^x_{A(B)}$ and $m^y_{A(B)}$, we find that in the symmetry-broken phase only the $x$ or $y$ component of the in-plane magnetization assumes a finite value, which gradually converges to a constant value as $J_{K}$ increases. The magnetization also turns out to be uniform, which means that it is independent of the sublattice degree of freedom [see Fig. \ref{MFSol_Qmin}(e)]. In the case of the self-consistent solution of the CMT order parameter shown in Fig. \ref{MFSol_Qmin}(f), we note that, despite the breaking of the spin symmetry of the conduction electrons in the ordered phase, its real $V_{R, \sigma}$ and imaginary $V_{I, \sigma}$ components with $\sigma \in \{ \uparrow, \downarrow\}$ remain equal to each other for all numerical mean-field solutions obtained in this work. We point out that these solutions were found without assuming any kind of constraint between $V_{R, \sigma}$ and $V_{I, \sigma}$. Finally, for electron dopings such that $|n_c| \lesssim 1$, our mean-field analysis suggests that the QPT out of the FL* phase turns out to be indeed discontinuous. 

\section{Instabilities of the conduction-electron system and ground-state phase diagram}\label{Sec_V}
\noindent

In order to describe the SC properties of the present model, we will examine the pairing self-energies of the conduction electrons, which are computed by integrating out the Majorana degrees of freedom of the Kitaev-Kondo lattice model. This procedure is equivalent to projecting the KSL Hamiltonian onto the Hilbert space of the complex fermions. As a consequence, the effective action of the conduction electrons becomes
\begin{align}\label{Eq_Eff_Action}
&\mathcal{S}_\text{eff} = \sum_{\omega_n}\sum_{\mathbf{k}\in(\text{BZ})/2}\begin{pmatrix} \Psi^\dagger_A(\mathbf{k},i\omega_n), & \Psi^\dagger_B(\mathbf{k},i\omega_n) \end{pmatrix}\nonumber\\
&\times[i\omega_n-\mathcal{H}_{t}(\mathbf{k})-\Sigma_K(\mathbf{k},i\omega_n)]\begin{pmatrix} \Psi_A(\mathbf{k},i\omega_n) \\ \Psi_B(\mathbf{k},i\omega_n) \end{pmatrix},
\end{align}
where
\begin{equation}\label{Eq_Balian}
\Psi_{A/B}(\mathbf{k},i\omega_n)=\begin{pmatrix} \psi_{A/B}(\mathbf{k},i\omega_n) \\ -i\sigma^y[\psi^\dagger_{A/B}(-\mathbf{k},-i\omega_n)]^T \end{pmatrix}
\end{equation}
is the Balian-Werthamer spinor in momentum-frequency space and $\Sigma_K(\mathbf{k},i\omega_n)$ is the fermionic self-energy obtained from the Kondo interaction between the KSL and the conduction electrons.

\begin{figure*}[t]
\centering 
\includegraphics[width=0.24\linewidth,valign=t]{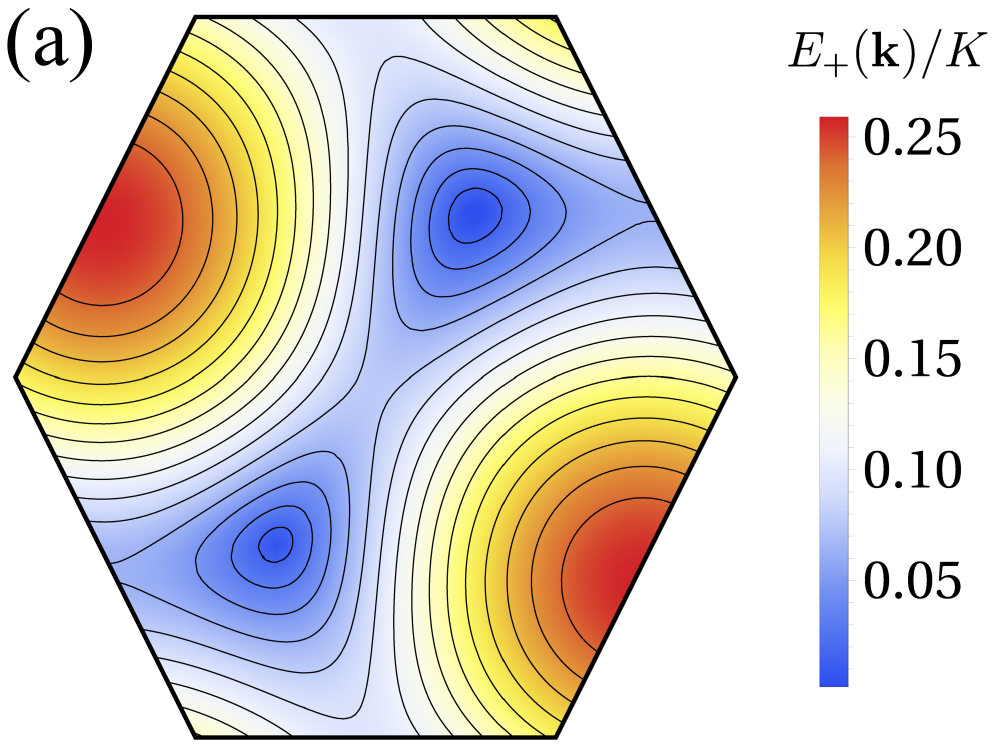} \hfil{}
\includegraphics[width=0.24\linewidth,valign=t]{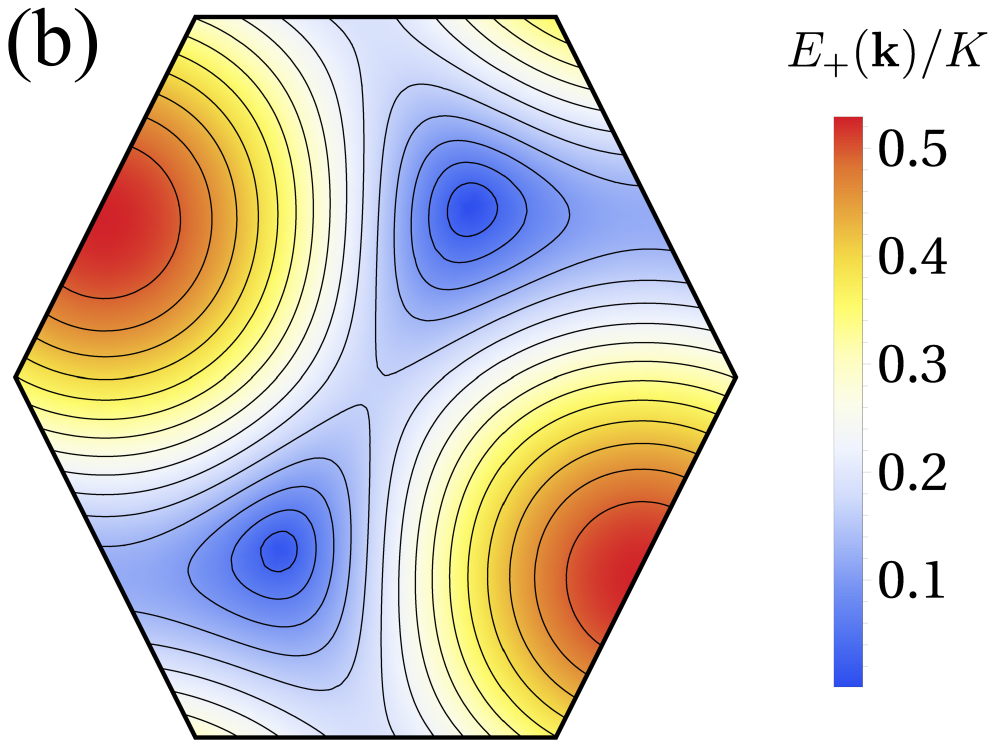} \hfil{}
\includegraphics[width=0.24\linewidth,valign=t]{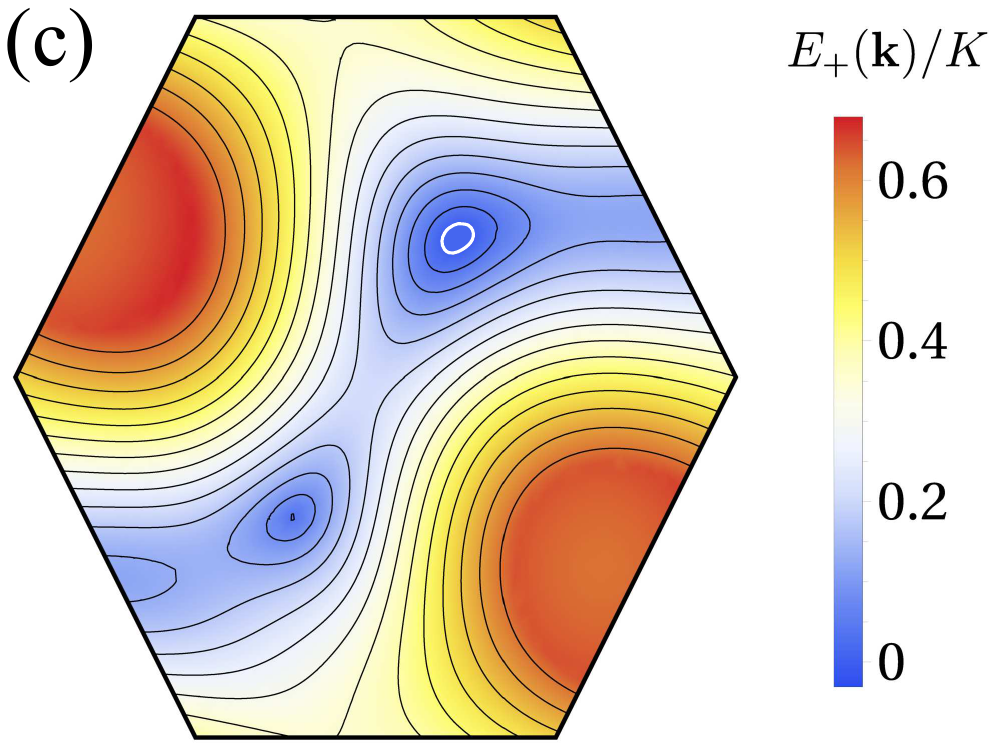} \hfil{}
\includegraphics[width=0.24\linewidth,valign=t]{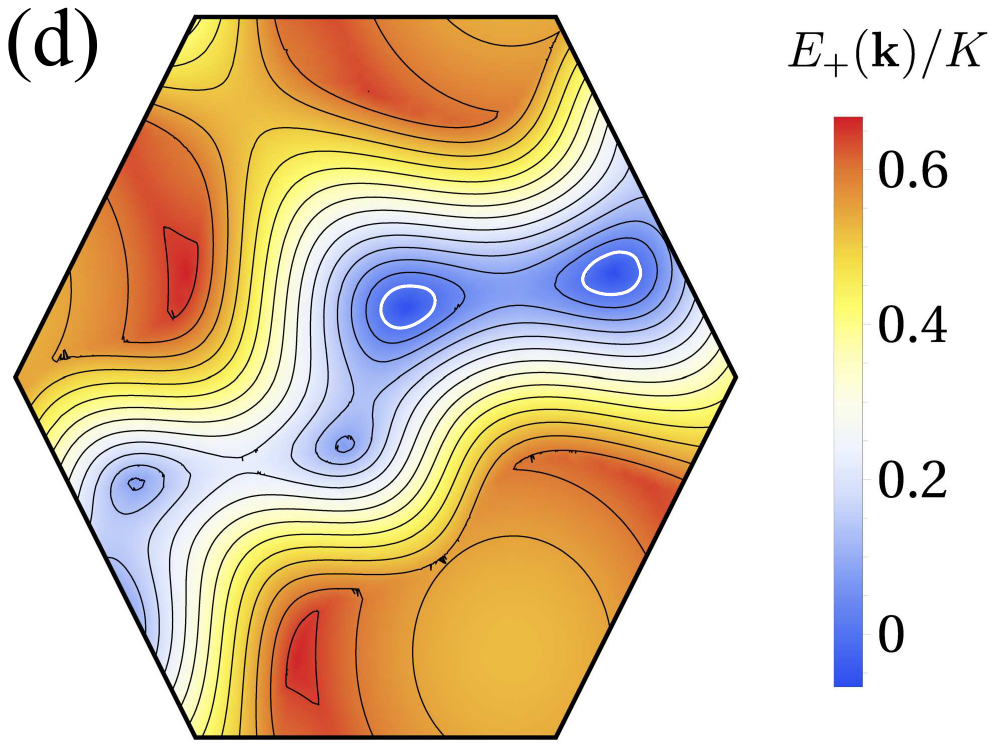} \vfil{}
\includegraphics[width=0.24\linewidth,valign=t]{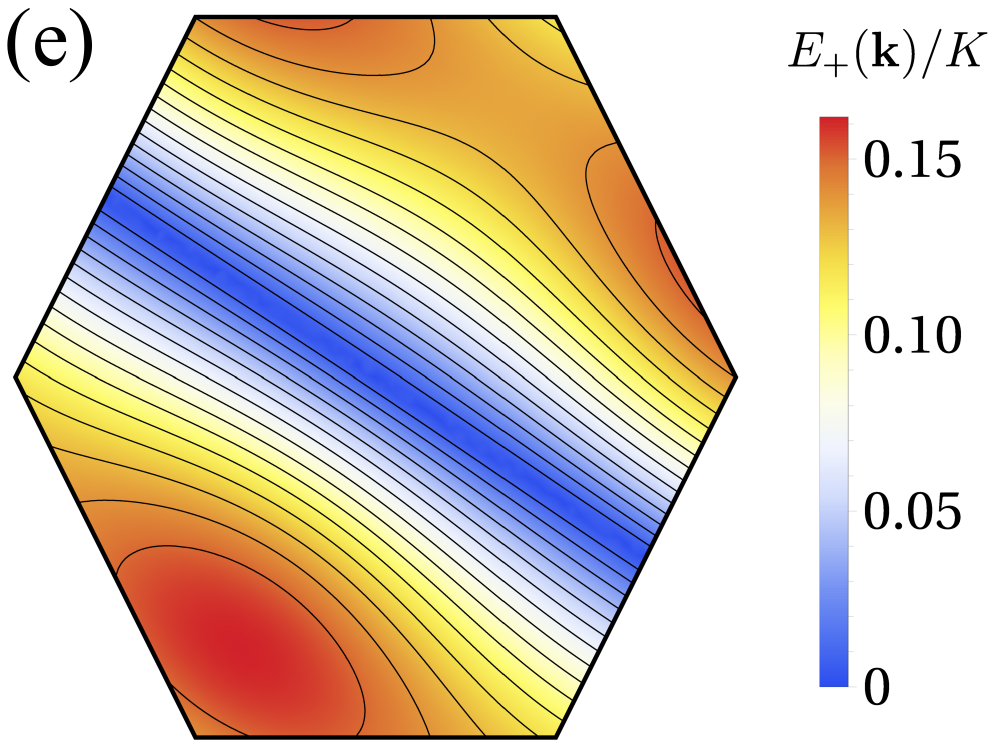} \hfil{}
\includegraphics[width=0.24\linewidth,valign=t]{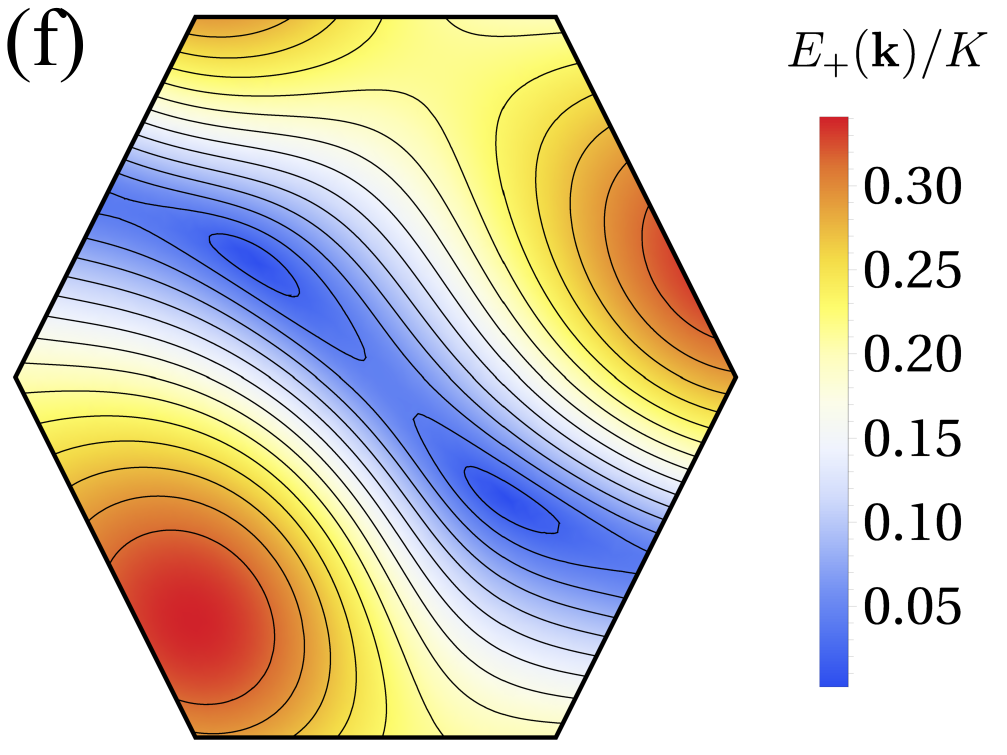} \hfil{}
\includegraphics[width=0.24\linewidth,valign=t]{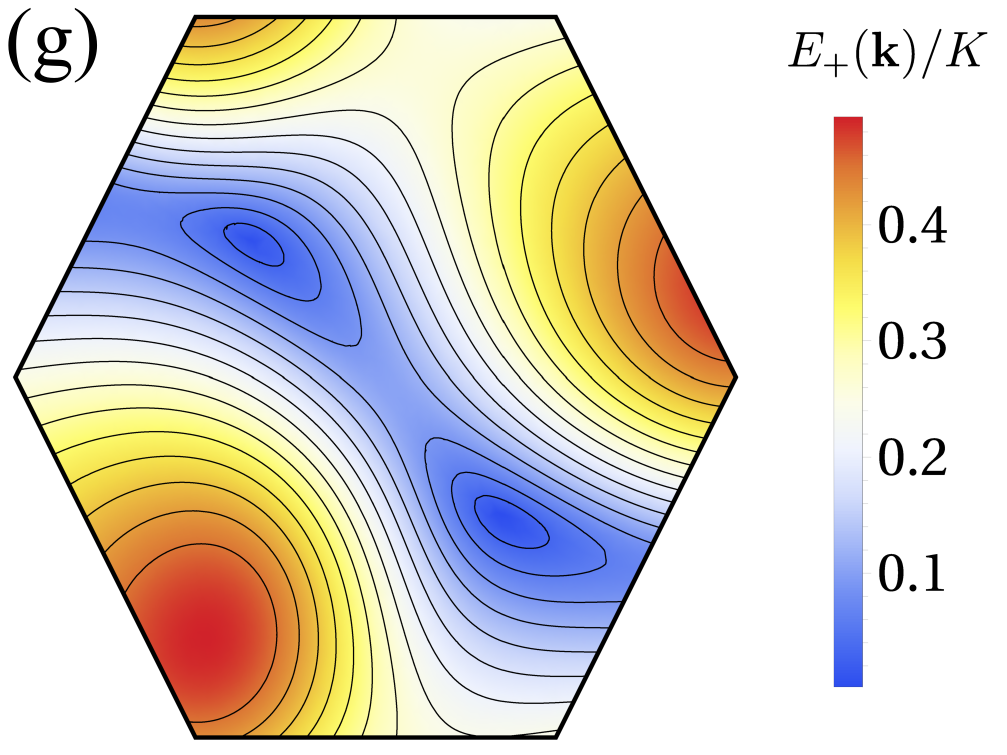} \hfil{}
\includegraphics[width=0.24\linewidth,valign=t]{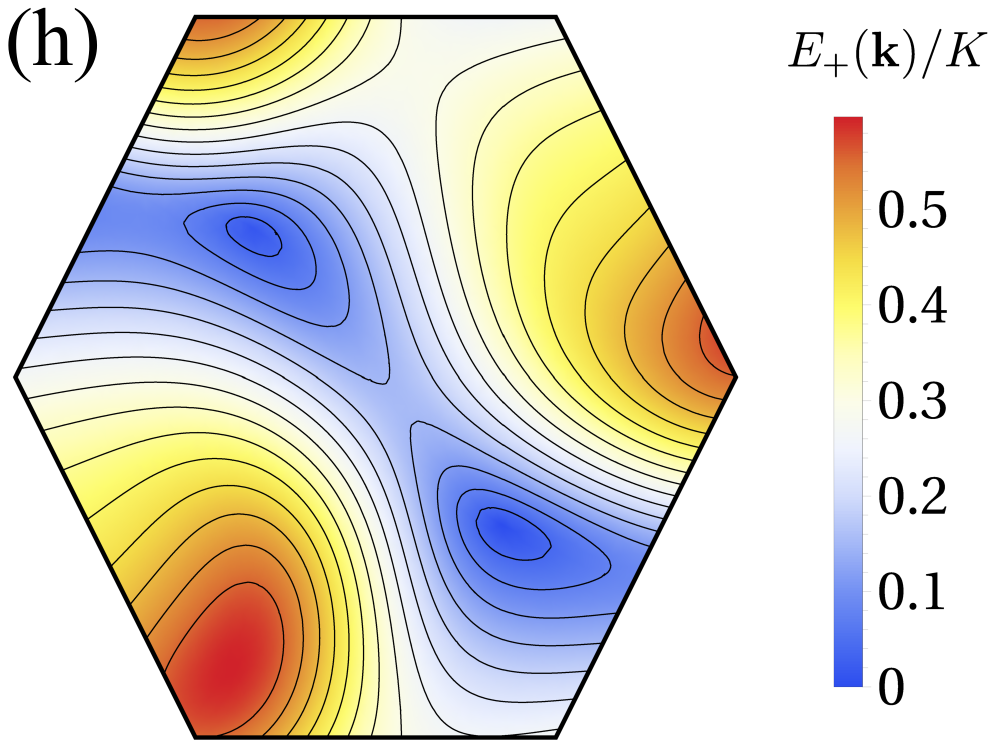}

\caption{Density plot within the first Brillouin zone of the band dispersion $E_{+}(\mathbf{k})$ closest to the zero of energy in the triplet SC phase of the Kitaev-Kondo lattice model for $J_K/K = 1.5$. In panels (a)--(d), we fix $n_c = 0.1$ and vary the electronic hopping as $t/K = 0.2, \; 0.4, \; 0.6$, and $0.8$. For $t/K \geq 0.6$, the Dirac nodal points transform into Bogoliubov-Fermi surfaces (represented by white closed lines). In panels (e)--(h), we fix $n_c = 0.9$ and increase the electronic hopping as before. As shown in panel (e), the nodal structure of the system for $t/K = 0.2$ is characterized by a Bogoliubov-Fermi line. As the hopping $t$ is increased, this Bogoliubov-Fermi line will evolve into two Dirac nodal points [panels (f)--(h)], while for even larger $t$ they will eventually transform into Bogoliubov-Fermi surfaces. As explained in the main text, the staggered phase $\mathbf{Q}$ is taken at the wave vector minimizing the mean-field ground state energy.}\label{BZ_Dens_Plots}
\end{figure*}

In order to evaluate the self-energy $\Sigma_K(\mathbf{k},i\omega_n)$ in the symmetry-broken phase, we use the numerical solution of the mean-field equations as an input, which is equivalent to setting $m^x_{A (B)} = m_x$, $m^y_{A (B)} = 0$, and $V_{R, \sigma} = V_{I, \sigma} = v$. One can easily show that $\Sigma_K(\mathbf{k},i\omega_n)$ splits into normal, resonant-exchange, and superconducting contributions \cite{Coleman-PRB(1994), Pereira-PRB(2020)}. As a result, we find that the mean-field effective action describing the SC fluctuations in the system evaluates to 
\begin{align}\label{Eff_SC_Action}
\mathcal{S}_\text{SC} = & \; \sum_{\mathbf{k}, \omega_n} \sum_{b b'} \psi_{b}^{T}(-\mathbf{k}, -i\omega_n)\Delta_\text{SC}^{b b'}(\mathbf{k}, i\omega_n) \psi_{b'}(\mathbf{k}, i\omega_n) \nonumber\\
& + \text{H.c.},
\end{align}
where $b \in \{A, B\}$ refers to the sublattice index, $\Delta_\text{SC}^{bb}(\mathbf{k}, i\omega_n) = i (v^2/2)\sigma_{y}(\boldsymbol{\sigma} \cdot \boldsymbol{D}^{b})$, and $\Delta_\text{SC}^{AB}(\mathbf{k}, i\omega_n) =  - \Delta_\text{SC}^{BA}(\mathbf{k}, i\omega_n) = i (v^2/2) \sigma_{y}(\boldsymbol{\sigma} \cdot \boldsymbol{D})$. In addition, the vectors in the expressions of the SC order parameters are in turn given by $\boldsymbol{D}^{b} \equiv D_{+}^{b}\hat{\mathbf{z}} - i D_{-}^{b}\hat{\mathbf{y}}$ and $\boldsymbol{D} \equiv D_{+}\hat{\mathbf{z}} - i D_{-}\hat{\mathbf{y}}$, where
\begin{widetext}
\begin{align}
    & D_{\pm}^{b} = \dfrac{2 Km_{x} \text{Im}[g_{y}(\mathbf{k})g_{z}^{*}(\mathbf{k})] (\delta_{B, b} - \delta_{A, b}) - \omega_{n}[ |g_{y}(\mathbf{k})|^{2} - |g_{z}(\mathbf{k})|^2]}{[\omega^2_n + f_2(\mathbf{k})/2]^2 + f_0(\mathbf{k}) - f^2_2(\mathbf{k})/4} \pm\dfrac{\omega_{n}}{\omega_{n}^{2}+\left|g_{x}(\mathbf{k})\right|^{2}}, \label{Gap_Eqs_01}\\
    & D_{\pm} = \dfrac{(\omega_{n}^{2}+K^{2}m_{x}^{2}) [g_{y}(\mathbf{k})-g_{z}(\mathbf{k})] + g_{y}(\mathbf{k}) |g_{z}(\mathbf{k})|^{2} - g_{z}(\mathbf{k}) |g_{y}(\mathbf{k})|^{2}}{[\omega^2_n + f_2(\mathbf{k})/2]^2 + f_0(\mathbf{k}) - f^2_2(\mathbf{k})/4} \pm \dfrac{g_{x}(\mathbf{k})}{\omega_{n}^{2}+\left|g_{x}(\mathbf{k})\right|^{2}},\label{Gap_Eqs_02}
\end{align}
\end{widetext}
where we have used the following definitions
\begin{align}
g_x(\mathbf{k}) \equiv & \; K u^{zy}e^{-i\mathbf{k}\cdot\mathbf{n}_2}, \\
g_y(\mathbf{k}) \equiv & \; K u^{zx}e^{-i\mathbf{k}\cdot\mathbf{n}_1}, \\
g_z(\mathbf{k}) \equiv & \; K (u^ye^{-i\mathbf{k}\cdot\mathbf{n}_1} + u^xe^{-i\mathbf{k}\cdot\mathbf{n}_2} + 1), \\
f_{0}(\mathbf{k}) \equiv & \; K^2 m^2_{x} [K^2 m^2_{x} - 2 \text{Re}( g_{y}(\mathbf{k})g^{*}_{z}(\mathbf{k}))] \nonumber \\
& + |g_{y}(\mathbf{k})|^{2} |g_{z}(\mathbf{k})|^{2}, \\
f_{2}(\mathbf{k}) \equiv & \; 2 K^{2}m_{x}^{2}+\left|g_{y}(\mathbf{k})\right|^{2}+\left|g_{z}(\mathbf{k})\right|^{2}.
\end{align}
Note that these results are for isotropic Kitaev interactions, i.e., $K_\gamma = K$. 

Note that the SC action in Eq. \eqref{Eq_Eff_Action} breaks inversion symmetry, since all the staggered phases $\mathbf{Q}$ found in the minimization of the ground-state energy do not represent time-reversal invariant wave vectors. Besides, this action describes a triplet state with both even- and odd-frequency pairing correlations. The former involves sites of both the same and different sublattices, while the latter refers only to sites of the same sublattice. This situation bears some resemblance with the exactly solvable model studied in Ref. \cite{Pereira-PRB(2020)}, in which an octopolar Kondo coupling between $j_\text{eff} = 3/2$ localized magnetic moments and conduction electrons of a non-centrosymmetric superconductor \cite{Mineev(2012), Yip-ARCMP(2014)} gives rise to a triplet pairing state with odd-frequency correlations for sites of the honeycomb lattice belonging to the same sublattice. However, in contrast to the model investigated in Ref. \cite{Pereira-PRB(2020)}, the SC state obtained here for the Kitaev-Kondo lattice model has one remarkable distinct feature: It exhibits a coexistence of both even- and odd-frequency pairing correlations for sites of the same sublattice, although the maximum value of the odd-frequency component of $\Delta_\text{SC}^{bb}(\mathbf{k}, i\omega_n)$ turns out to be always the dominant contribution (for more details, see Appendix \ref{Appendix_B}). We point out that the simultaneous presence of odd- and even-frequency pairing amplitudes is fundamentally related to the underlying time-reversal symmetry breaking in the system that mixes both components. Indeed, one can see that the even-frequency component clearly vanishes if the magnetization is set to zero (i.e., with the restoration of time-reversal symmetry). Moreover, the validity of the Fermi-Dirac statistics for the SC correlations is always satisfied here because the even-frequency component of $\Delta_\text{SC}^{bb}(\mathbf{k}, i\omega_n)$ is odd under parity, while the odd-frequency one is even under the same operation.

\begin{figure}[t]
\centering
\includegraphics[width=1.0\linewidth,valign=t]{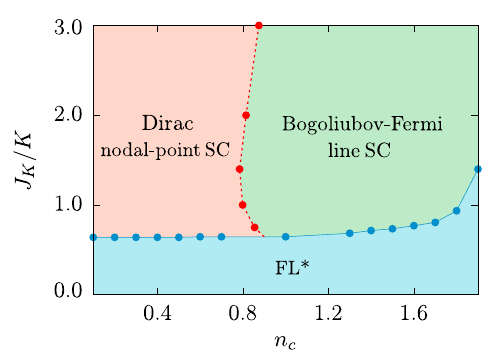}
\caption{Ground-state phase diagram of the Kitaev-Kondo lattice model as a function of the Kondo interaction $J_K$ and the conduction band filling $n_c$. The electronic hopping is fixed here to $t/K = 0.2$. The main features of this phase diagram are the emergence of both FL* and a triplet SC phase. As explained in the main text, the SC order possesses even-frequency and odd-frequency pairing correlations and can exhibit for $t \ll K$ either Dirac points or Bogoliubov-Fermi lines as nodal structures of the bulk spectrum. When $t \gtrsim K$, these structures evolve into Bogoliubov-Fermi surfaces when the system enters the SC state. The staggered phase $\mathbf{Q}$ is taken at the wave vector minimizing the mean-field ground state energy.}\label{PDiagram}
\end{figure}

In Fig. \ref{BZ_Dens_Plots}, we show the dependence of the energy dispersion $E_{+}(\mathbf{k})$ closest to the zero-energy level, which is obtained inside the symmetry-broken SC phase of the present model. Depending on the conduction band filling $n_c$ and the hopping $t$ of the model, the SC phase can exhibit either Dirac points, Bogoliubov-Fermi lines or Bogoliubov-Fermi surfaces \cite{Brydon-PRB(2018)} as nodal manifolds of the bulk spectrum. In fact, as shown in Figs. \ref{BZ_Dens_Plots}(a)--\ref{BZ_Dens_Plots}(d), the Dirac nodes appear in the SC state when the band filling $n_c$ and the hopping $t$ are both small. As the hopping is increased in this case, the Dirac nodes evolve into Bogoliubov-Fermi surfaces. On the other hand, when the hopping is fixed for $t \ll K$ and the band filling $n_c$ is raised, the Dirac nodes of $E_{+}(\mathbf{k})$ turn into a nodal Bogoliubov-Fermi line, which crosses the entire Brillouin zone and disperses linearly along its transversal direction [see Fig. \ref{BZ_Dens_Plots}(e)]. As shown in Figs. \ref{BZ_Dens_Plots}(f)--\ref{BZ_Dens_Plots}(h), this Bogoliubov-Fermi line can evolve again into Dirac nodes by simply setting the electronic hopping to $t \sim K$. As the parameter $t$ is further increased, the Dirac nodes will become eventually Bogoliubov-Fermi surfaces, as before. In addition, it is also evident from Fig. \ref{BZ_Dens_Plots} that these SC states break the $C_3$ point group symmetry of the original FL* system, giving rise to a nematic superconductor. Indeed, this can be explicitly confirmed by the evaluation of the nearest-neighbor SC correlations along each one of the three types of bonds of the honeycomb lattice. 

\begin{figure}[b]
\centering 
\includegraphics[width=1.0\linewidth,valign=t]{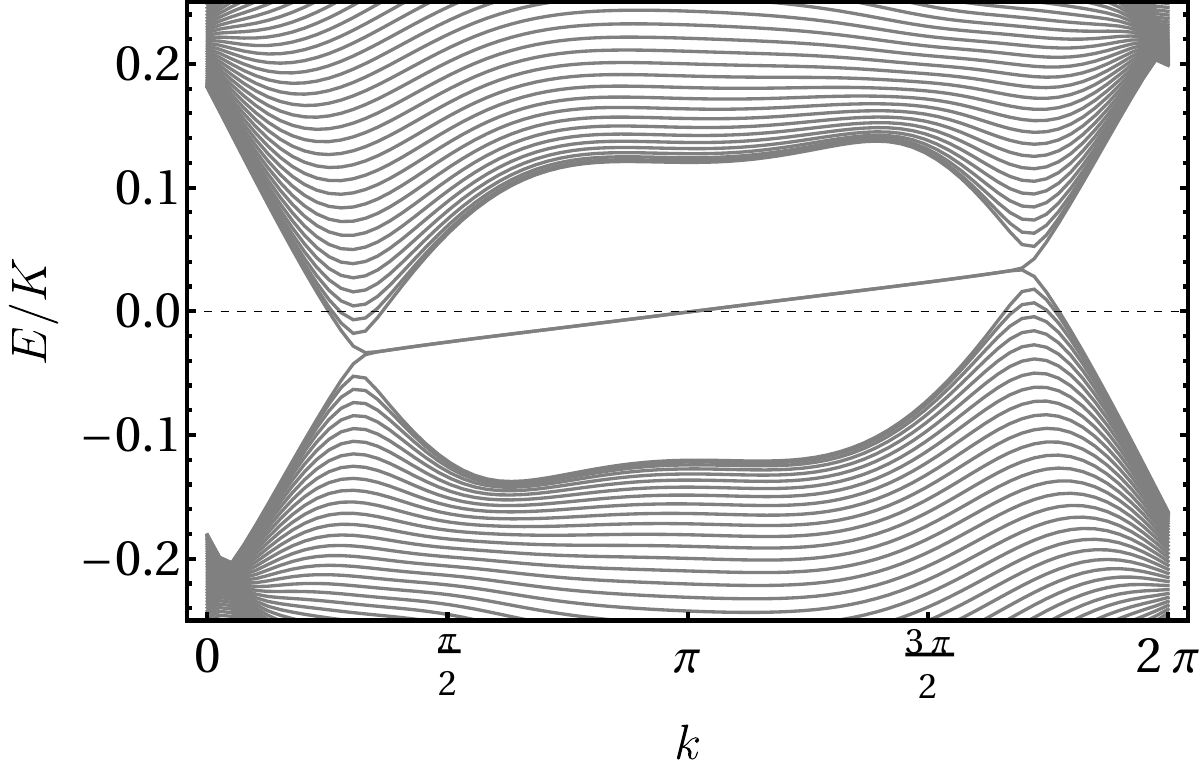}
\caption{Surface-state dispersions in the SC phase with a Bogoliubov-Fermi surface for a finite system with zigzag edge along the $\hat{\mathbf{x}}$ direction. In this case, the surface states display the presence of antichiral modes, which co-propagate in the same direction along both parallel edges in a strip geometry. The model parameters are fixed here as $J_K/K = 1.5$, $n_c = 0.1$, and $t/K = 0.6$. The staggered phase $\mathbf{Q}$ is taken at the wave vector minimizing the mean-field ground state energy.}\label{Surface_Dispersions}
\end{figure}

In Fig. \ref{PDiagram}, we show the mean-field phase diagram of the Kitaev-Kondo lattice model at $T \rightarrow 0$ as a function of the Kondo interaction $J_K$ and the conduction electron band filling $n_c$, when the electronic hopping is fixed to $t/K = 0.2$. We focus here only on the interval of doping defined by $0 < n_c < 2$, since the other half of the phase diagram (i.e. for densities satisfying $2 < n_c < 4$) turns out to be a simple reflection of the first part due to the particle-hole symmetry with respect to half-filling (i.e., $n_c = 2$). As shown in Fig. \ref{PDiagram}, we obtain that the Kitaev-Kondo model indeed displays a QPT from a FL* at small $J_K$ to a SC phase with triplet pairing at larger $J_K$ for all values of $n_c$. Also according to Fig. \ref{PDiagram} and already anticipated from Fig. \ref{BZ_Dens_Plots}, the phase diagram at small values of the hopping $t$ exhibits an abrupt Lifshitz transition \cite{Lifshitz-JETP(1959), Volovik-LTP(2017)} (i.e., a transition associated with a change of the nodal structure) between the gapless SC states with the Dirac nodes and the Bogoliubov-Fermi line. As explained before, the further increase of $t$ changes somewhat this phase diagram, such that both the Dirac nodes and the Bogoliubov-Fermi line observed in the SC phase transform into fully developed Bogoliubov-Fermi surfaces. Remarkably, the SC phase associated with the latter nodal manifolds is also characterized by topologically protected surface states having gapless antichiral modes \cite{Franz-PRL(2018)}, which have the distinct feature of co-propagating in the same direction along both parallel edges in a strip geometry (see Fig. \ref{Surface_Dispersions}), with the gapless bulk modes formed by the Bogoliubov quasi-particles supplying the necessary counter-propagating modes in the system. 

Finally, we need to comment on how our results are connected to those obtained in Refs. \cite{Vojta-PRB(2018),Kim-PRB(2018)}. As explained earlier, those previous works also discussed the phase diagram of the Kitaev-Kondo lattice model on the bilayer honeycomb lattice using different mean-field theories. In fact, their calculations start from a representation of the spin-$1/2$ magnetic moments in terms of complex (Abrikosov) fermions. By applying this formalism, both Refs. \cite{Vojta-PRB(2018),Kim-PRB(2018)} found that the Kitaev-Kondo lattice model undergoes a first-order QPT from an FL* to a SC state with triplet pairing, which is also in qualitative agreement with our present results. However, the main difference between those two previous works is that the SC state obtained in Ref. \cite{Vojta-PRB(2018)} is always gapless and topologically trivial, while in Ref. \cite{Kim-PRB(2018)} the authors obtain fully gapped SC phases as a function of $J_K$, which turn out to be topologically non-trivial. This difference stems from the way the mean-field decoupling of both KSL and Kondo Hamiltonians is performed.  

\begin{figure}[t]
\includegraphics[width=0.45\textwidth,valign=t]{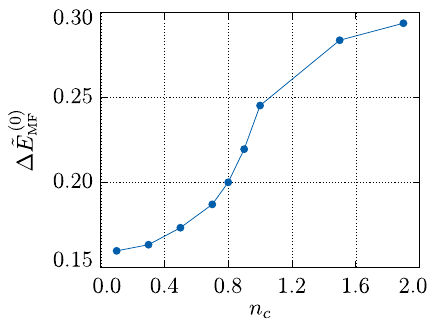}
\caption{Dimensionless energy difference $\Delta \tilde{E}^{(0)}_{MF} \equiv [E^{(0)}_{MF}(\mathbf{Q=0})-E^{(0)}_{MF}(\mathbf{Q=Q^*})]/K$ between the SC ground states associated with the wave vector $\mathbf{Q}=0$ and the non-zero staggered wave vector $\mathbf{Q^*}$ for the corresponding hybridization order parameter, as a function of the doping $n_c$ and for fixed $t/K = 0.2$. The order parameter with wave vector $\mathbf{Q^*}$ stabilizes the mean-field ground state, which is always associated with the lowest ground-state energy in the model.}
\label{GS_Energy_Diff} 
\end{figure}

Furthermore, those previous works that analyzed the Kitaev-Kondo lattice model assumed that the SC order parameter of the model is even in frequency and non-zero only between conduction electrons localized on sites of different sublattices. This should be contrasted with the approach presented in this work, where we obtain both odd-frequency and even-frequency components for the SC order parameter. In addition, we also consider that the hybridization order parameter of the conduction electrons with the spin-$1/2$ moments -- used to decouple the Kondo interaction -- acquires a non-zero staggered wave vector $\mathbf{Q^*}$ in order to stabilize the mean-field ground state. This staggered SC state is associated with an energy lower than the $\mathbf{Q}=0$ state (this is shown, e.g., in Fig. \ref{GS_Energy_Diff}, where we compare the ground-state energies in the present model as a function of the doping $n_c$). Since the finite staggering phase $\mathbf{Q^*}$ impacts on the SC phase obtained here as a finite center-of-mass momentum of the Cooper pairs, this phase turns out to be in fact a PDW state in the present system.

Lastly, we comment on the the fact that the mean-field theory of the Kitaev-Kondo lattice model in terms of the SO(3) Majorana representation has one limitation regarding the regime where the Kitaev-Kondo interaction approaches the limit $t \sim J_K \gg K$. The reason for this is because the mean-field solution found in this case cannot describe the heavy Fermi-liquid phase of the model \cite{Vojta-PRB(2018)}. Despite this cautionary remark, we argue here that the SO(3) Majorana representation is expected to provide a more accurate description (compared to other mean-field approaches) of the SC phase in the Kitaev-Kondo lattice model in the regime $t \ll J_K \sim K$, since it takes into account both the frequency dependence and the staggering nature of the SC order parameter. In this sense, the present SO(3) Majorana mean-field approach can be viewed as a complementary approach for describing an important part of the phase diagram of this system.

\section{Concluding remarks}\label{Sec_VI}

In summary, we have studied the Kitaev-Kondo lattice model on the honeycomb lattice by starting from an SO(3) Majorana representation for the localized spin-$1/2$ magnetic moments to decouple the Kitaev interaction and at the same time employing the CMT order parameter, which involves a combination of a complex and a Majorana fermion, to decouple the Kondo interaction. This approach was found to give rise to both even- and odd-frequency pairing components between the conduction electrons on the honeycomb lattice. In addition, the SC states that emerge in this case have triplet pairing correlations, nematic order, and break time-reversal symmetry. We have also pointed out that the SC phases obtained here are in fact PDW states that are accompanied by ferromagnetic order, which persists for larger values of electron doping and Kondo interaction.

We have shown that the QPT out of the FL* is towards a gapless SC phase having either Dirac nodes, Bogoliubov-Fermi line, or Bogoliubov-Fermi surfaces. The two former states occur for lower values of the electron hopping, while the latter takes place when this quantity becomes comparable to the Kitaev interaction. This result was compared with other works that analyzed the properties of the Kitaev-Kondo lattice model, which were based on the application of mean-field theory in terms of Abrikosov fermions to represent the spin-$1/2$ magnetic moments \cite{Vojta-PRB(2018), Kim-PRB(2018)}. We have emphasized that the present approach takes into account the full dependence of the hybridization (i.e, the CMT order parameter) between the conduction electrons and the Majorana fermions of the KSL on the staggered wave vector $\mathbf{Q}$. As a consequence, we have found new ground-states in the model that minimize the corresponding energy with respect to this wave vector. Finally, we hope that the present evidence of PDW states with odd-frequency pairing component will motivate future research on potential candidate materials that may be described by the Kitaev-Kondo lattice model.

\begin{acknowledgments}

We thank D. Chakraborty and R. G. Pereira for important discussions and C. de Farias for her collaboration in the early stages of this work. V.S.deC. acknowledges the financial support from CAPES through a post-doctoral fellowship No. 88887.469170/2019-00. R.M.P.T. thanks a Ph.D. fellowship from CAPES. Financial support from CNPq is acknowledged by H.F. (310710/2018-9) and E.M. (307041/2017-4). E.M. also acknowledges Capes/Cofecub 0899/2018.

\end{acknowledgments}

\appendix

\section{Mean-field gap equations}\label{Appendix_A}
\noindent

The mean-field Hamiltonian for the Kitaev-Kondo lattice can be diagonalized by the unitary transformation
\begin{equation}\label{Eq_Unit_Transf}
\Upsilon(\mathbf{k}) = \mathcal{U}(\mathbf{k})\chi(\mathbf{k}),
\end{equation}
where $\Upsilon(\mathbf{k}) = (\Psi^T_A(\mathbf{k}), \Psi^T_B(\mathbf{k}), c^x_A(\mathbf{k}), c^x_B(\mathbf{k}), \ldots, c^z_B(\mathbf{k}))^T$ is a $14$-component spinor. The substitution of this last expression into the mean-field Kitaev-Kondo Hamiltonian yields
\begin{align}
H_\text{\tiny MF} & = \sum_{\mathbf{k}}\sum^{14}_{\ell=1}E_\ell(\mathbf{k})\chi^\dagger_\ell(\mathbf{k})\chi_\ell(\mathbf{k}) - \mathcal{N} K_x u^y u^{zx}  \nonumber \\
& - \mathcal{N} K_y u^x u^{zy} + \mathcal{N} K_x m^x_A m^x_B + \mathcal{N} K_y m^y_A m^y_B  \nonumber \\
& + 4 \mathcal{N} \frac{( V^2_R + V^2_I )}{J_K}, \label{Eq_Diag_MF_Hamiltonian}
\end{align}
where $\mathcal{N}$ refers to the number of lattice unit cell, $V^2_R \equiv V^2_{R, \uparrow} + V^2_{R, \downarrow}$, $V^2_I \equiv V^2_{I, \uparrow} + V^2_{I, \downarrow}$, and $E_\ell(\mathbf{k})$ are the energy dispersions which, due to the complexity of the Hamiltonian, have to be evaluated numerically.

In view of the expressions in Eqs. \eqref{Eq_Unit_Transf} and \eqref{Eq_Diag_MF_Hamiltonian}, the mean-field equations for the bond variables for finite Kondo interaction evaluate to
\begin{align}
u^{x (y)} = & \; \int_\text{BZ} \frac{d^2 \mathbf{k}}{\mathcal{A}_\text{BZ}} \sum^{14}_{\ell=1}\big[\mathcal{U}^\dagger(\mathbf{k})\Lambda_{x (y)}(\mathbf{k})\mathcal{U}(\mathbf{k})\big]_{\ell,\ell} \nonumber \\
& \times n_F[E_\ell(\mathbf{k})],\\
u^{zx (zy)} = & \; \int_\text{BZ} \frac{d^2 \mathbf{k}}{\mathcal{A}_\text{BZ}} \sum^{14}_{\ell=1}\big[\mathcal{U}^\dagger(\mathbf{k})\Lambda_{zx (zy)}(\mathbf{k})\mathcal{U}(\mathbf{k})\big]_{\ell,\ell} \nonumber \\
& \times n_F[E_\ell(\mathbf{k})],\\
m^{x(y)}_{A (B)} = & \; - \int_\text{BZ} \frac{d^2 \mathbf{k}}{\mathcal{A}_\text{BZ}} \sum^{14}_{\ell=1}\big[\mathcal{U}^\dagger(\mathbf{k})\mathcal{M}^{x(y)}_{A (B)}(\mathbf{k})\mathcal{U}(\mathbf{k})\big]_{\ell,\ell} \nonumber \\
& \times n_F[E_\ell(\mathbf{k})],
\end{align}
where $\mathcal{A}_\text{BZ} = 8 \pi^2/(3 \sqrt{3} a^2)$ is the area of the Brillouin zone and $n_F(x) \equiv (e^{\beta x} + 1)^{- 1}$ is the Fermi-Dirac distribution function. In addition, in order to write down the above equations, we have employed the following matrices
\begin{align}
\Lambda_x(\mathbf{k}) & =
\begin{pmatrix}
\mathbb{0}_{8\times8} & \mathbb{0}_{8\times2} & \mathbb{0}_{8\times2} & \mathbb{0}_{8\times2} \\ 
\mathbb{0}_{2\times8} & \begin{pmatrix} 0 & ie^{-i\mathbf{k}\cdot\mathbf{n}_2} \\ -ie^{i\mathbf{k}\cdot\mathbf{n}_2} & 0 \end{pmatrix} & \mathbb{0}_{2\times2} & \mathbb{0}_{2\times2} \\
\mathbb{0}_{2\times8} & \mathbb{0}_{2\times2} & \mathbb{0}_{2\times2} & \mathbb{0}_{2\times2} \\
\mathbb{0}_{2\times8} & \mathbb{0}_{2\times2} & \mathbb{0}_{2\times2} & \mathbb{0}_{2\times2} 
\end{pmatrix}, \\
\Lambda_y(\mathbf{k}) & =
\begin{pmatrix}
\mathbb{0}_{8\times8} & \mathbb{0}_{8\times2} & \mathbb{0}_{8\times2} & \mathbb{0}_{8\times2} \\ 
\mathbb{0}_{2\times8} & \mathbb{0}_{2\times2} & \mathbb{0}_{2\times2} & \mathbb{0}_{2\times2} \\
\mathbb{0}_{2\times8} & \mathbb{0}_{2\times2} & \begin{pmatrix} 0 & ie^{-i\mathbf{k}\cdot\mathbf{n}_1} \\ -ie^{i\mathbf{k}\cdot\mathbf{n}_1} & 0 \end{pmatrix} & \mathbb{0}_{2\times2} \\
\mathbb{0}_{2\times8} & \mathbb{0}_{2\times2} & \mathbb{0}_{2\times2} & \mathbb{0}_{2\times2}
\end{pmatrix}, \\
\Lambda_{zx}(\mathbf{k}) & =
\begin{pmatrix}
\mathbb{0}_{8\times8} & \mathbb{0}_{8\times2} & \mathbb{0}_{8\times2} & \mathbb{0}_{8\times2} \\ 
\mathbb{0}_{2\times8} & \mathbb{0}_{2\times2} & \mathbb{0}_{2\times2} & \mathbb{0}_{2\times2} \\
\mathbb{0}_{2\times8} & \mathbb{0}_{2\times2} & \mathbb{0}_{2\times2} & \mathbb{0}_{2\times2} \\
\mathbb{0}_{2\times8} & \mathbb{0}_{2\times2} & \mathbb{0}_{2\times2} & \begin{pmatrix} 0 & ie^{-i\mathbf{k}\cdot\mathbf{n}_1} \\ -ie^{i\mathbf{k}\cdot\mathbf{n}_1} & 0 \end{pmatrix} 
\end{pmatrix}, \\
\Lambda_{zy}(\mathbf{k}) & =
\begin{pmatrix}
\mathbb{0}_{8\times8} & \mathbb{0}_{8\times2} & \mathbb{0}_{8\times2} & \mathbb{0}_{8\times2} \\ 
\mathbb{0}_{2\times8} & \mathbb{0}_{2\times2} & \mathbb{0}_{2\times2} & \mathbb{0}_{2\times2} \\
\mathbb{0}_{2\times8} & \mathbb{0}_{2\times2} & \mathbb{0}_{2\times2} & \mathbb{0}_{2\times2} \\
\mathbb{0}_{2\times8} & \mathbb{0}_{2\times2} & \mathbb{0}_{2\times2} & \begin{pmatrix} 0 & ie^{-i\mathbf{k}\cdot\mathbf{n}_2} \\ -ie^{i\mathbf{k}\cdot\mathbf{n}_2} & 0 \end{pmatrix} 
\end{pmatrix},
\end{align}
\begin{align}
\mathcal{M}^x_A(\mathbf{k}) & = 
\begin{pmatrix} \mathbb{0}_{10 \times 2} & \mathbb{0}_{10 \times 2} & \mathbb{0}_{10 \times 2} \\
\mathbb{0}_{2 \times 2} & \mathbb{0}_{2 \times 2} & \begin{pmatrix} i & 0 \\ 0 & 0 \end{pmatrix} \\
\mathbb{0}_{2 \times 2} & \begin{pmatrix} - i & 0 \\ 0 & 0 \end{pmatrix} & \mathbb{0}_{2 \times 2}
\end{pmatrix}, \\
\mathcal{M}^x_B(\mathbf{k}) & = 
\begin{pmatrix} \mathbb{0}_{10 \times 2} & \mathbb{0}_{10 \times 2} & \mathbb{0}_{10 \times 2} \\
\mathbb{0}_{2 \times 2} & \mathbb{0}_{2 \times 2} & \begin{pmatrix} 0 & 0 \\ 0 & i \end{pmatrix} \\
\mathbb{0}_{2 \times 2} & \begin{pmatrix} 0 & 0 \\ 0 & - i \end{pmatrix} & \mathbb{0}_{2 \times 2}
\end{pmatrix}, \\
\mathcal{M}^y_A(\mathbf{k}) & = 
\begin{pmatrix} \mathbb{0}_{2 \times 10} & \mathbb{0}_{2 \times 2} & \begin{pmatrix} - i & 0 \\ 0 & 0 \end{pmatrix} \\
\mathbb{0}_{10 \times 2} & \mathbb{0}_{10 \times 2} & \mathbb{0}_{10 \times 2} \\
\begin{pmatrix} i & 0 \\ 0 & 0 \end{pmatrix} & \mathbb{0}_{2 \times 10} & \mathbb{0}_{2 \times 2}
\end{pmatrix}, \\
\mathcal{M}^y_B(\mathbf{k}) & = 
\begin{pmatrix} \mathbb{0}_{2 \times 10} & \mathbb{0}_{2 \times 2} & \begin{pmatrix} 0 & 0 \\ 0 & - i \end{pmatrix} \\
\mathbb{0}_{10 \times 2} & \mathbb{0}_{10 \times 2} & \mathbb{0}_{10 \times 2} \\
\begin{pmatrix} 0 & 0 \\ 0 & i \end{pmatrix} & \mathbb{0}_{2 \times 10} & \mathbb{0}_{2 \times 2}
\end{pmatrix}.
\end{align}

\begin{figure*}[t]
\centering 
\includegraphics[width=0.45\linewidth,valign=t]{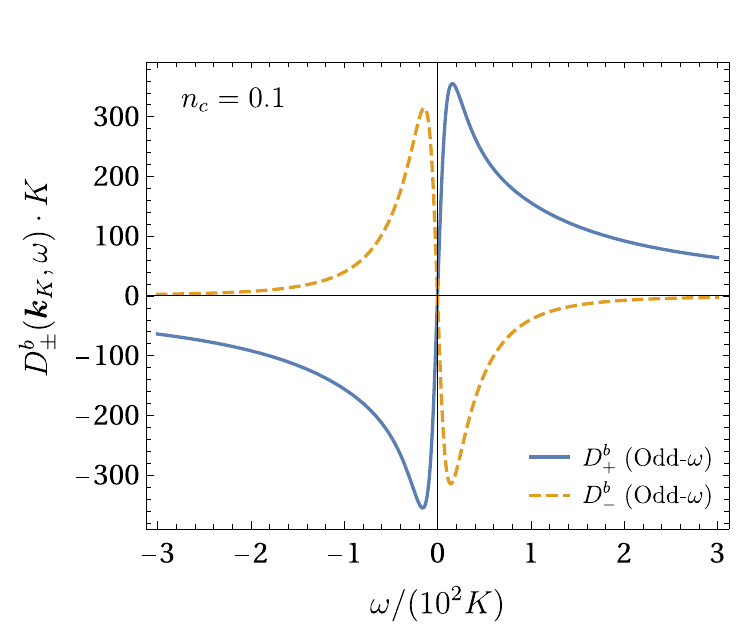} \hfil{}
\includegraphics[width=0.45\linewidth,valign=t]{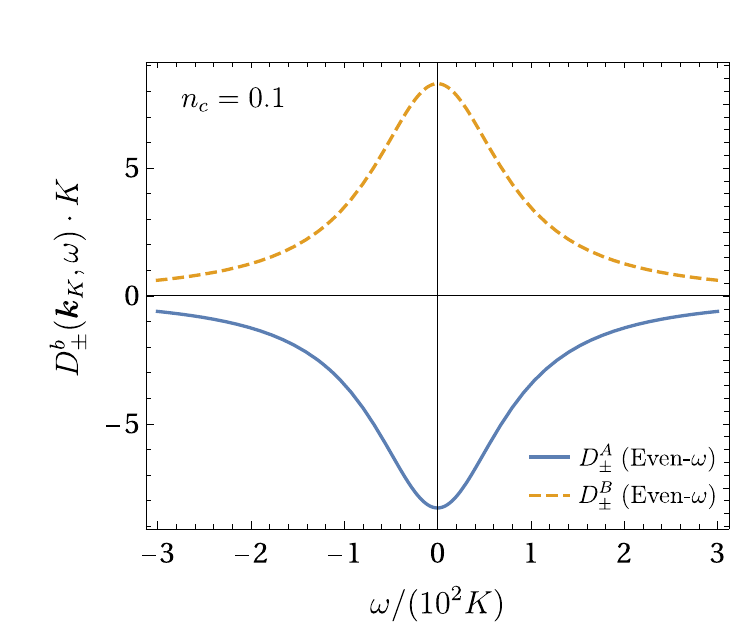} \vfil{}
\includegraphics[width=0.45\linewidth,valign=t]{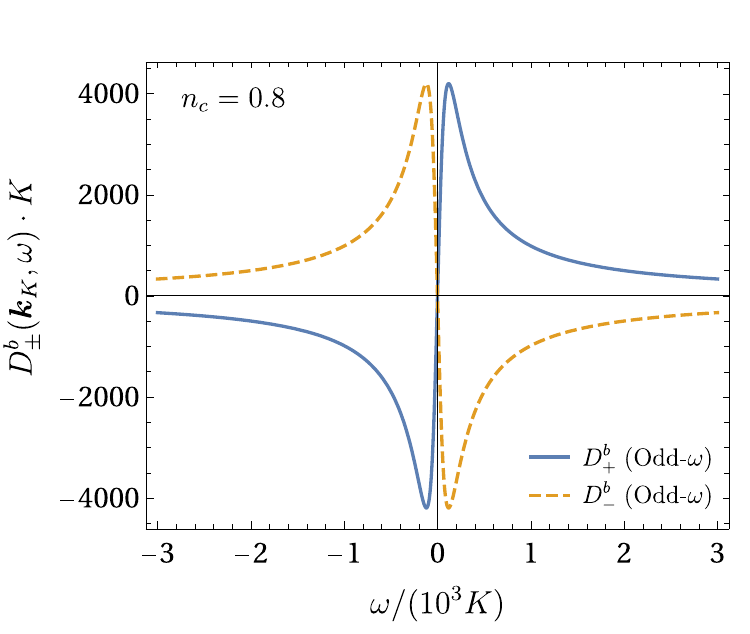} \hfil{}
\includegraphics[width=0.45\linewidth,valign=t]{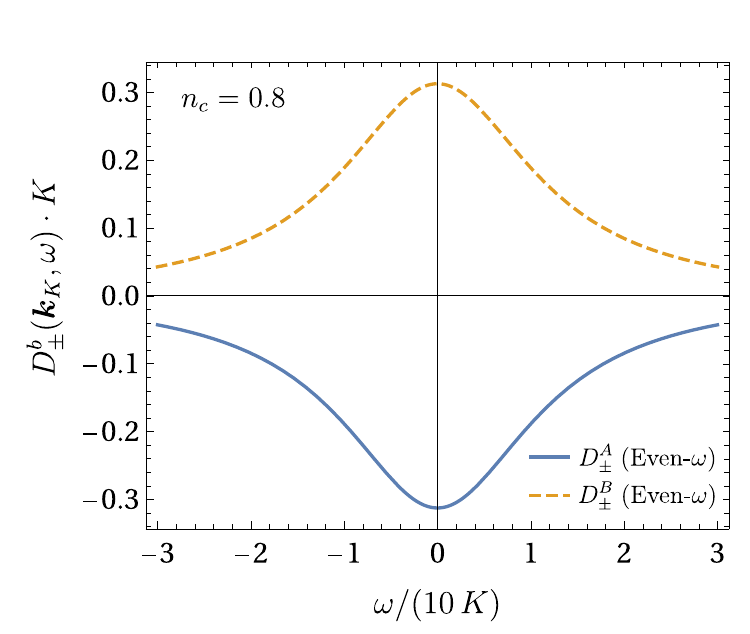}
\caption{The $\omega$-dependence of the odd- and even-frequency components of the SC correlation function $D^b_\pm(\mathbf{k}, \omega)$ for the wavevector $\mathbf{k} = \mathbf{K}$, as the electronic filling varies from $n_c = 0.1$ (upper panels) to $n_c = 0.8$ (lower panels).}\label{Even_Odd_SC_Gap}
\end{figure*}

On the other hand, the mean-field equations for the components of the CMT order parameter are given by
\begin{align}
V_{R/I, \sigma} = & - \frac{J_K}{8} \int_\text{BZ} \frac{d^2 \mathbf{k}}{\mathcal{A}_\text{BZ}} \sum^{14}_{\ell=1}\big[\mathcal{U}^\dagger(\mathbf{k})\Lambda_{R/I, \sigma}(\mathbf{k})\mathcal{U}(\mathbf{k})\big]_{\ell,\ell} \nonumber \\
& \times n_F[E_\ell(\mathbf{k})],
\end{align}
where $\Lambda_{R/I, \sigma}(\mathbf{k})=
\begin{pmatrix}
\mathbb{0}_{8\times8} & \varSigma_{R/I, \sigma} \\
\varSigma^\dagger_{R/I, \sigma} & \mathbb{0}_{6\times6}
\end{pmatrix}.
$ Besides, the off-diagonal block matrix $\varSigma_{R/I, \sigma}$ refers to
\begin{widetext}
\begin{equation}
\varSigma_{R/I, \sigma} = \frac{1}{\sqrt{2}} \begin{pmatrix} (\sigma^x\otimes\mathbb{1})\mathcal{I}_{R/I, \sigma}\!\!\! & \mathbb{0}_{4\times1}\!\!\! & (\sigma^y\otimes\mathbb{1})\mathcal{I}_{R/I, \sigma}\!\!\! & \mathbb{0}_{4\times1}\!\!\! & (\sigma^z\otimes\mathbb{1})\mathcal{I}_{R/I, \sigma}\!\!\! & \mathbb{0}_{4\times1}\!\!\! \\ \mathbb{0}_{4\times1}\!\!\! & (\sigma^x\otimes\mathbb{1})\mathcal{I}_{R/I, \sigma}\!\!\! & \mathbb{0}_{4\times1}\!\!\! & (\sigma^y\otimes\mathbb{1})\mathcal{I}_{R/I, \sigma}\!\!\! & \mathbb{0}_{4\times1}\!\!\! & (\sigma^z\otimes\mathbb{1})\mathcal{I}_{R/I, \sigma}\end{pmatrix},
\end{equation}
\end{widetext}
\begin{equation}
\mathcal{I}_{R, \uparrow} = \begin{pmatrix} 1 \\ 0 \\ 0 \\ 1 \end{pmatrix}, \; \mathcal{I}_{R, \downarrow} = \begin{pmatrix} 0 \\ 1 \\ -1 \\ 0 \end{pmatrix}, 
\end{equation}
\begin{equation}
 \mathcal{I}_{I, \uparrow} = \begin{pmatrix} i \\ 0 \\ 0 \\ -i \end{pmatrix}, \; \mathcal{I}_{I, \downarrow} = \begin{pmatrix} 0 \\ i \\ i \\ 0 \end{pmatrix}.
\end{equation}

Furthermore, Eqs. \eqref{Eq_Unit_Transf} and \eqref{Eq_Diag_MF_Hamiltonian} also allow us to write down in a compact manner the expression for the conduction-electron density per unit cell. Indeed, we obtain
\begin{equation}
n_c=\frac{1}{2} \int_\text{BZ} \frac{d^2 \mathbf{k}}{\mathcal{A}_\text{BZ}} \sum^{14}_{\ell=1}\big[\mathcal{U}^\dagger(\mathbf{k})\Lambda_c(\mathbf{k})\mathcal{U}(\mathbf{k})\big]_{\ell,\ell}n_F[E_\ell(\mathbf{k})],
\end{equation}
where
\begin{equation}
\Lambda_c(\mathbf{k})=
    \begin{pmatrix}
    \begin{pmatrix} T^z & \mathbb{0}_{4\times4} \\ \mathbb{0}_{4\times4} & T^z \end{pmatrix} & \mathbb{0}_{8\times6} & \\ 
    \mathbb{0}_{6\times8} & \mathbb{0}_{6\times6} 
    \end{pmatrix}.
\end{equation}

Finally, the ground-state energy of the model per unit cell is given by
\begin{align}
E^{\text{\tiny (0)}}_\text{\tiny MF} = & \int_\text{BZ} \frac{d^2 \mathbf{k}}{\mathcal{A}_\text{BZ}} \sum^{14}_{\ell=1}E_\ell(\mathbf{k})\Theta[- E_\ell(\mathbf{k})] - K_x u^y u^{zx}  \nonumber \\
& - K_y u^x u^{zy} + K_x m^x_A m^x_B + K_y m^y_A m^y_B  \nonumber \\
& + 4 \frac{( V^2_R + V^2_I )}{J_K},
\end{align}
with $\Theta (x)$ being the Heaviside step function. The physical staggered wave vector $\mathbf{Q}$ of the CMT order parameter is the one that minimizes $E^{\text{\tiny (0)}}_\text{\tiny MF}$.

\section{Behavior of the even- and odd-frequency components of $D^b_\pm(\mathbf{k}, \omega)$}\label{Appendix_B}
\noindent

In Fig. \ref{Even_Odd_SC_Gap}, we plot the even- and odd-frequency components of SC correlation function $D^b_\pm(\mathbf{k}, \omega)$ defined in Eq. \eqref{Gap_Eqs_01} for the wavevector $\mathbf{k} = \mathbf{K}$ at the corner of the Brillouin zone. Note that the maximum values of the odd-frequency components of $D^b_\pm(\mathbf{k}, \omega)$ always represent the largest contributions of this correlation function. We have also checked numerically that this behavior does not alter, as the momentum $\mathbf{k}$ and the electronic filling $n_c$ are changed.

%\bibliographystyle{apsrev4-1_control}
%\bibliography{./Superconductors_Ref}

%merlin.mbs apsrev4-1.bst 2010-07-25 4.21a (PWD, AO, DPC) hacked
%Control: key (0)
%Control: author (72) initials jnrlst
%Control: editor formatted (1) identically to author
%Control: production of article title (0) allowed
%Control: page (0) single
%Control: year (1) truncated
%Control: production of eprint (0) enabled
%

\end{document}